\documentclass[journal]{IEEEtran}
\usepackage{etex}
\usepackage{cite}
\ifCLASSINFOpdf
   \usepackage[pdftex]{graphicx}
  \graphicspath{{Journal_Figs/}{/}}
  \DeclareGraphicsExtensions{.pdf}
\else
   \usepackage[dvips]{graphicx}
\fi

\usepackage[cmex10]{amsmath}
\usepackage{amssymb}
\usepackage{upgreek}
\newtheorem{theorem}{Theorem}

\newtheorem{definition}{Definition}
\newtheorem{remark}{Remark}
\newtheorem{claim}{Claim}
\newtheorem{proposition}{Proposition}

\usepackage{algorithm}
\usepackage{algpseudocode}
\algrenewcommand\algorithmicrequire{\textbf{Intialization:}}
\algrenewcommand\algorithmicensure{\textbf{Output:}}

\usepackage{array}
\usepackage{multirow}
\usepackage{rotating}
\usepackage[font=footnotesize]{subfig}
\usepackage{fixltx2e}
\usepackage{mathtools, cuted}
\usepackage{float}
\usepackage{url}
\DeclareMathOperator{\Ss}{\texttt{S}}
\DeclareMathOperator{\R}{\texttt{R}}
\DeclareMathOperator{\D}{\texttt{D}}
\DeclareMathOperator{\E}{\texttt{E}}
\DeclareMathOperator{\eve}{\texttt{eve}}

\DeclareMathOperator{\tr}{Tr}
\DeclareMathOperator{\vect}{vec}
\DeclareMathOperator{\SINR}{\texttt{SINR}}
\DeclareMathOperator{\st}{\text{s.t.}}
\DeclareMathOperator{\rea}{\mathrm{Re}}

\DeclareMathOperator{\rank}{\mathrm{Rank}}

\DeclareMathOperator{\blkdiag}{\texttt{blkdiag}}
\DeclareMathOperator{\K}{\mathcal{K}}
\DeclareMathOperator{\x}{\mathbf{x}}
\DeclareMathOperator{\xn}{\mathbf{x}^{(\mathit{n})}}
\DeclareMathOperator{\xnn}{\mathbf{x}^{(\mathit{n}+1)}}
\DeclareMathOperator{\G}{\pmb{\mathcal{G}}}
\DeclareMathOperator{\F}{\pmb{\mathcal{F}}}
\DeclareMathOperator{\algone}{\text{Algorithm~1}}

\begin{document}
\title{Secure MIMO Relaying Network: An Artificial Noise Aided Robust Design Approach}

\author{Jiaxin Yang,~\IEEEmembership{Student Member,~IEEE,} Qiang Li,~\IEEEmembership{Member, IEEE,} Benoit Champagne,~\IEEEmembership{Senior Member,~IEEE,}\\Yulong Zou,~\IEEEmembership{Senior Member,~IEEE,} and Lajos Hanzo,~\IEEEmembership{Fellow,~IEEE}
        % <-this % stops a space
\thanks{J.~Yang and B.~Champagne are with the Department
of Electrical and Computer Engineering, McGill University, Montreal,
Quebec, H3A~0E9, Canada. (E-mail: jiaxin.yang@mail.mcgill.ca; benoit.champagne@mcgill.ca).

Q.~Li is with the School of Communication and Information Engineering, University of Electronic Science and Technology of China, Chengdu, 611731, China. (E-mail: lq@uestc.edu.cn).

Y.~Zou is with the School of Telecommunications and Information Engineering, Nanjing University of Posts and Telecommunications, Nanjing, 210003, P. R. China. (E-mail: yulong.zou@njupt.edu.cn).

L.~Hanzo is with the School of Electronics and Computer Science, University of Southampton, Southampton, SO17~1BJ, U.K.. (E-mail: lh@ecs.soton.ac.uk).}% <-this % stops a space
% <-this % stops a space
}

\maketitle

\begin{abstract}
%\boldmath
	Owing to the vulnerability of relay-assisted and device-to-device (D2D) communications, improving wireless security from a physical layer signal processing perspective is attracting increasing interest. Hence we address the problem of secure transmission in a relay-assisted network, where a pair of legitimate user equipments (UEs) communicate with the aid of a multiple-input multiple output (MIMO) relay in the presence of multiple eavesdroppers ($\eve$s). Assuming imperfect knowledge of the $\eve$s' channels, we jointly optimize the power of the source UE, the amplify-and-forward (AF) relaying matrix and the covariance of the artificial noise (AN) transmitted by the relay, in order to maximize the received signal-to-interference-plus-noise ratio (SINR) at the destination, while imposing a set of \emph{robust secrecy constraints}. To tackle the resultant non-convex optimization problem, a globally optimal solution based on a bi-level optimization framework is proposed, but with high complexity. Then a low-complexity sub-optimal method relying on a new penalized difference-of-convex (DC) algorithmic framework is proposed, which is specifically designed for non-convex semidefinite programs (SDPs). We show how this penalized DC framework can be invoked for solving our robust secure relaying problem with proven convergence. Our extensive simulation results show that both proposed solutions are capable of ensuring the secrecy of the relay-aided transmission and significantly improve the robustness towards the $\eve$s' channel uncertainties as compared to the non-robust counterparts. It is also demonstrated the penalized DC-based method advocated yields a performance close to the globally optimal solution.
\end{abstract}


% \fi
\IEEEpeerreviewmaketitle

\section{Introduction}
With the proliferation of smartphones storing more sensitive personal data ranging from social networking to online banking, wireless end-users have become vulnerable targets of hackers. According to a recent report on mobile cyber threats, the number of cyber attacks to mobile users has been dramatically growing, e.g., by nearly 10-fold from August 2013 to March 2014 \cite{Mobile_threats}. Within this context, how to ensure information security is becoming a critical issue for wireless service providers. Although the classic bit-level encryption technique has been deemed to be most effective way of achieving this goal, a recent report by the Washington Post has drawn public attention to the potential security risks of wireless technologies, even when advanced encryption is used\footnote{In \cite{WP}, it is reported that two German researchers have demonstrated how to exploit the security flaws in the Signaling System 7 (SS7) to eavesdrop on all incoming and outgoing calls indefinitely from anywhere in the world. They have shown how to decode the messages by requesting each caller's carrier to release a temporary encryption key through the SS7.}.
Against this background, physical layer security is emerging as a promising design alternative to complement classic encryption and to further enhance the security of wireless networks.

Since Wyner opened this new avenue of security provision by introducing the notion of secrecy capacity \cite{wyner}, researchers have sought to enhance security for a wide range of communication channel models, as discussed in \cite{Information_security,phy_security_multiantenna,phy_security_survey} and the references therein. Recently, physical layer security has attracted increased interest, driven by new techniques such as cooperative relaying, which has found its way into the Long-Term Evolution (LTE) standard. Although the diversity advantages gleaned from user cooperation have been recognized in the context of generic relay-assisted networks \cite{Unified,xing}, ensuring secrecy in message relaying remains a key issue. Specifically, when additional intermediate nodes assist in forwarding the source messages, the information confidentiality may be more readily compromised, unless the relaying scheme is appropriately designed. It was demonstrated in \cite{Lai} that relaying is capable of improving the level of security. This seminal work has led to further research endeavors devoted to investigating the secrecy of relay-assisted communications from the physical layer perspective \cite{Secrecy_Cooperation}. Following this trend, in this paper emphasis will be on new signal processing techniques conceived for improving wireless relaying security. Below we briefly review related works on this research topic and summarize our main contributions.

\subsection{Related Works}
A wireless relay can adopt either the amplify-and-forward (AF) or the decode-and-forward (DF) strategy for forwarding source messages. For DF relaying, the optimal weights that achieve the maximum secrecy capacity are derived in \cite{Relay_Security_AFDF,Relay_Security_DF}. As compared to DF, AF relaying offers its inherent advantages of lower signal processing complexity and latency, and hence will be the focus of our attention. A variety of relaying solutions such as beamforming, cooperative jamming and artificial noise (AN) generation, or a hybrid of the aforementioned options, have been studied in \cite{Relay_Beamforming,6823730,6051523,6655008,6956810,Robust_Secure_Relaying,Robust_AF_AN,6819004,6560019,Ding,Joint}. For instance, the optimal AF relaying weights maximizing the achievable secrecy rate of a single-antenna relay network are derived in \cite{Relay_Beamforming}, without consideration of the source information leakage to $\eve$s. When multiple antennas are employed at both the source and relay, joint transmit precoding and power allocation relying on the generalized singular value decomposition (GSVD) is proposed in \cite{6823730}.
Finally, joint source transmit precoding and multi-antenna AF relaying is investigated in \cite{6051523} assuming an untrusted relay node.

The contributions \cite{Relay_Beamforming,6823730,6051523} assume perfect knowledge of each $\eve$'s channel state information (ECSI) at the legitimate nodes. In practice, due to the lack of explicit cooperation between the latter and $\eve$s, at best an inaccurate estimate of the ECSI may be available\footnote{A notable example is the device-to-device (D2D) discovery and communication defined in 3GPP LTE Rel. 12 \cite{D2D}. Each UE (including the potential $\eve$s) periodically broadcasts its own beacon signals and listens to others using a small portion of the LTE uplink resources. In this way, each UE is able to discover the presence of other UEs (including potential $\eve$s in its proximity) and subsequently infers an imprecise ECSI estimate based on the approximate distance and pathloss coefficients.}. Assuming that the ECSI errors lie in a predefined norm-bounded region, joint relay beamforming and jamming signal design in a single-antenna relay network is developed in \cite{6655008,6956810} with the objective of maximizing the worst-case secrecy rate. Extension of this approach to a more generalized model where multi-antenna is employed at the relay is considered in  \cite{Robust_Secure_Relaying,Robust_AF_AN}. Minimization of the mean square error (MSE) of the received signal at the destination, subject to a set of signal-to-interference-plus-noise (SINR)-based secrecy constraints, is considered in \cite{6819004}. Using the same uncertainty model, the problem of total relaying power minimization is studied in \cite{6560019,Ding} by simultaneously guaranteeing a predefined quality-of-service (QoS) level at the destination and a certain secrecy level against eavesdropping. Finally, \cite{Robust_AF_AN} assumes a more general relay system configuration, where some of the prior works can be viewed as a special case. In this work, a globally optimal solution is obtained resorting to a bi-level optimization framework, where the upper-level problem is tackled by one-dimensional search, while the inner-level problem is solved by semidefinite relaxation (SDR) \cite{SDR}.

\subsection{Contributions}
This paper considers a general wireless communication scenario, where a source ($\Ss$) transmits its confidential data to a destination ($\D$), assisted by a multi-antenna AF relay ($\R$). Both phases of the two-hop transmission are overheard by a set of independent $\eve$s. The power of $\Ss$, the AF relaying matrix and the covariance matrix of the AN emitted by $\R$ have to be jointly optimized for protecting the message confidentiality. In contrast to most of the prior contributions \cite{Relay_Security_AFDF,Relay_Security_DF,Relay_Beamforming,6823730,6655008,6956810,Robust_Secure_Relaying,Robust_AF_AN}, where the main focus has been on the maximization of the (worst-case) secrecy rate when either perfect or imperfect ECSI is available, we consider the problem from the alternative perspective of the security-reliability trade-off recently introduced in \cite{SRT,Yulong_Survey}. Specifically, assuming that the ECSI errors reside in a spherical region, we aim for maximizing the received SINR at $\D$, subject to specific power constraints, while satisfying a set of \emph{robust secrecy constraints} at $\eve$s. Our contributions are detailed as follows:
\begin{itemize}
	\item We find the global optimum for the formulated non-convex problem by reformulating the latter as a bi-level optimization problem, where an SDR-based solution is derived for the inner problem and the tightness of such a relaxation is proved.
	\item We propose a new penalized difference-of-convex (DC) algorithmic framework for a class of general non-convex semidefinite programs (SDPs), which eliminates the need for a non-trivial feasible initialization routinely required by the conventional DC algorithm \cite{CCP}. We explicitly prove that the solution sequence generated by the algorithm converges to a stationary point of the original problem.
	%\item Furthermore, we propose an inexact version of the penalized DC framework imposing a reduced-complexity, which allows the sub-problem at each iteration to be solved inexactly as opposed to the exact penalized DC algorithm. We explicitly point out the required conditions to guarantee convergence.
	\item We show how the secrecy-constrained robust relaying problem can be transformed into the form solvable by the penalized DC algorithm and subsequently, we apply the proposed algorithm to solve the design problem efficiently.
	\item We perform extensive numerical simulations for comparing the performance of the proposed solutions to other benchmarking schemes. We demonstrate that the penalized DC-based solution is capable of achieving approximately the same performance as the globally optimal solution at a significantly reduced complexity.
\end{itemize}

\subsection{Organization and Notations}
The rest of the paper is organized as follows. Section II introduces the relay system model and formulates our secrecy-constrained robust relaying problem. The globally optimal solution relying on the SDR and on a one-dimensional exhaustive search is presented in Section III. In Section IV, we propose a penalized DC algorithmic framework and characterize its convergence. We then invoke the proposed framework for solving our secure relaying problem in Section V. The performance of the proposed solutions is quantified via numerical simulations in Section VI. Finally, we conclude in Section VII.

Boldface uppercase (lowercase) letters denote matrices (vectors), while normal letters denote scalars; $(\cdot)^*$, $(\cdot)^T$, $(\cdot)^H$, and $(\cdot)^{-1}$ denote the conjugate, transpose, Hermitian transpose and inverse, respectively; $\lVert\cdot\rVert$ represents the Euclidean norm of a vector, while $\lVert\cdot\rVert_F$ denotes the Frobenius norm of a matrix; $\tr(\cdot)$, $\vect(\cdot)$, and $\otimes$ stand for the matrix trace, vectorization and the Kronecker product, respectively; $\mathbb{C}^{M\times M}$ and $\mathbb{H}^M$ denotes the spaces of $M\times M$ matrices having complex entries and $M\times M$ Hermitian matrices, respectively; $\rea\{\cdot\}$ denotes the real part of a complex number; $\lambda_{\max}[\cdot]$ denotes the largest eigenvalue of a Hermitian matrix.



\begin{figure}[t]
	\centering
	\includegraphics[width=6.5cm]{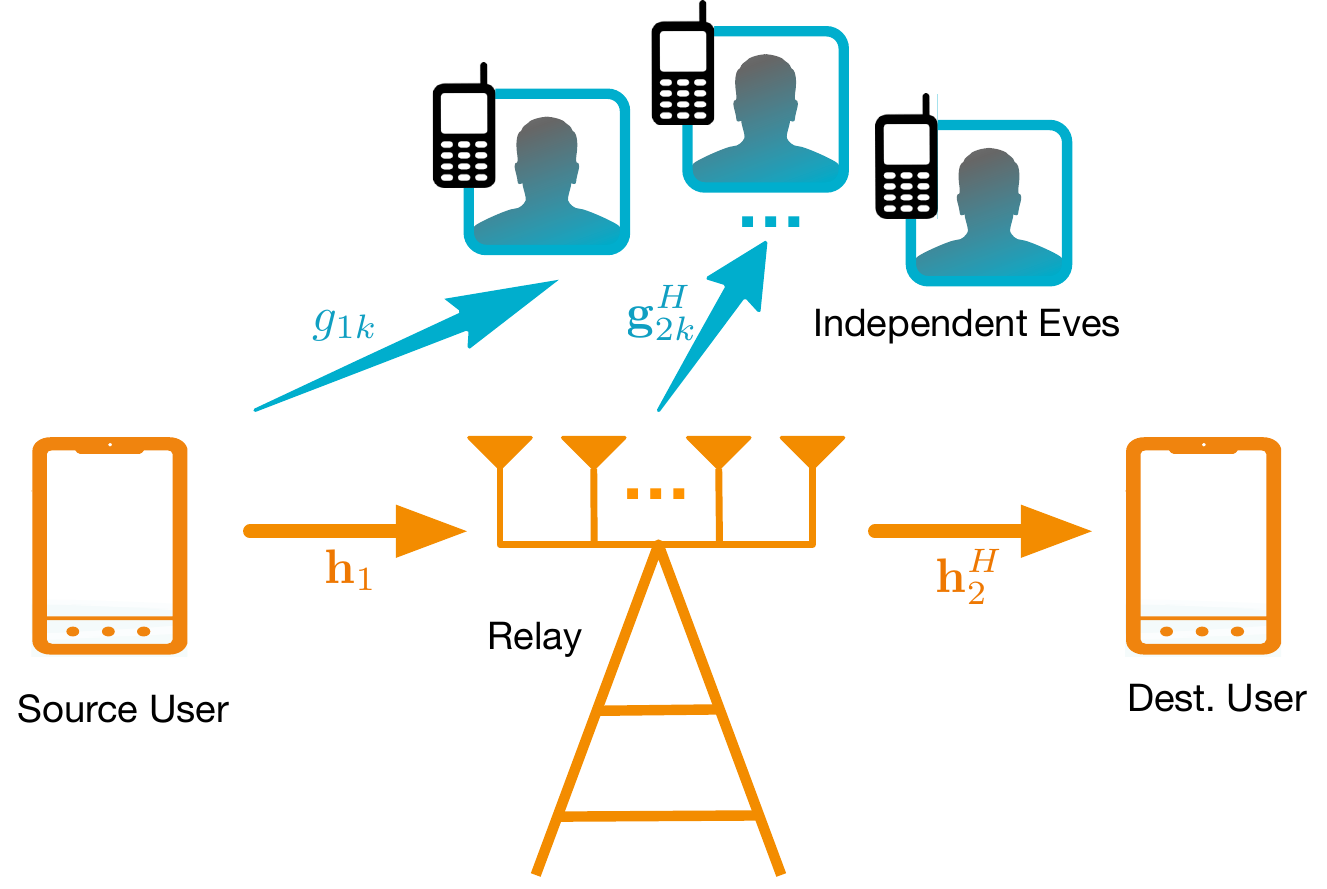}\\
	\caption{MIMO relay network in the presence of multiple single-antenna $\eve$s.}\label{fig1}\vspace{-10pt}
\end{figure}
\section{System Model and Problem Formulation}\label{sec2}
We commence by presenting the specific communication scenario, where multiple independent $\eve$s can potentially overhear the relay-assisted transmission, and subsequently formulate the robust secure relaying problem.
\subsection{MIMO Relay System in the Presence of Eavesdroppers}
Consider the wireless network as depicted in Fig.~\ref{fig1}, where source $\Ss$ communicates with destination $\D$, assisted by a trusted AF relay $\R$ operating in a half-duplex mode. The signals transmitted during the $\Ss\rightarrow\R$ and $\R\rightarrow\D$ hops are overheard by $K$ independent $\eve$s, $\E_k$ for $k\in\K\triangleq\left\{1,2,\cdots,K\right\}$. We assume that $\Ss$, $\D$ and $\E_k,~\forall k\in\K$ are single-antenna UEs having limited signal processing capabilities and low power budgets. By contrast, $\R$ is equipped with $N_{\R}\geq 2$ antennas. It is assumed that no direct link is available between $\Ss$--$\D$ due to the severe pathloss.

A narrowband flat-fading channel model is considered, where we denote the $\Ss$--$\R$ channel by $\mathbf{h}_1\in\mathbb{C}^{N_{\R}\times1}$ and the Hermitian transpose of the $\R$--$\D$ channel by $\mathbf{h}_2\in\mathbb{C}^{N_{\R}\times1}$. Let $s$ denote the $\Ss$ information symbol, modeled as a zero-mean Gaussian random variable with a power of $\sigma_{\Ss}^2\leq P_{\Ss}$, where $P_{\Ss}$ denotes the $\Ss$ power budget. During the first transmission slot, the signal received at $\R$ is given by
\begin{equation}\label{sec2_1}
\mathbf{z}=\mathbf{h}_1s+\mathbf{n}_{\R},
\end{equation}
where $\mathbf{n}_{\R}$ is a zero-mean additive noise vector with covariance of $\sigma_{\R}^2\mathbf{I}_{N_{\R}}$.
Then $\R$ applies a linear AF transformation matrix $\mathbf{W}\in\mathbb{C}^{N_{\R}\times N_{\R}}$ to the received signal, and superimposes an AN vector onto the linearly processed signal. Hence, the signal to be forwarded to \texttt{D} is given by
\begin{equation}\label{sec2_2}
\mathbf{r}=\mathbf{W}\mathbf{z}+\mathbf{v}=\mathbf{W}\mathbf{h}_1s+\mathbf{W}\mathbf{n}_{\R}+\mathbf{v},
\end{equation}
where $\mathbf{v}$ denotes the AN vector with zero mean and covariance of $\mathbb{E}\{\mathbf{v}\mathbf{v}^H\}=\pmb{\Psi}$ to be optimized.
The relay $\R$ has the power constraint of
$\sigma_{\Ss}^2\lVert\mathbf{W}\mathbf{h}_1\rVert^2+\sigma_{\R}^2\lVert\mathbf{W}\rVert_F^2+\tr(\pmb{\Psi})\leq P_{\R}$,
where $P_{\R}$ denotes its power budget. During the second transmission slot, $\D$ receives the following signal:
\begin{equation}\label{sec2_3}
	y_{\D}=\mathbf{h}_2^H\mathbf{W}\mathbf{h}_1s+\mathbf{h}_2^H\mathbf{W}\mathbf{n}_{\R}+\mathbf{h}_2^H\mathbf{v}+n_{\D},
\end{equation}
where $n_{\D}$ is an additive noise with zero mean and a variance of $\sigma_{\D}^2$.

We adopt, as a metric of transmission reliability, the received SINR at $\D$ given by
\begin{equation}\label{sec2_4}
\SINR_{\D}=\frac{\sigma_{\Ss}^2|\mathbf{h}_2^H\mathbf{W}\mathbf{h}_1|^2}{\sigma_{\R}^2\lVert\mathbf{h}_2^H\mathbf{W}\rVert^2+\mathbf{h}_2^H\pmb{\Psi}\mathbf{h}_2+\sigma_{\D}^2}.
\end{equation}
During the transmission, each $\E_k$ is potentially capable of overhearing the signals transmitted both from $\Ss$ and $\R$. Let $g_{1k}$ and $\mathbf{g}_{2k}\in\mathbb{C}^{N_{\R}\times 1}$, respectively, denote the $\Ss$--$\E_k$ channel and the Hermitian transpose of the $\R$--$\E_k$ channel. Then the signals observed by $\E_k$ from $\Ss$ and $\R$, respectively, are given by
\begin{eqnarray}
\label{sec2_5}
y_{\E,k}^{\Ss}&=& g_{1k}s+n_{\E,1k}\\
\label{sec2_6}
y_{\E,k}^{\R}&=& \mathbf{g}_{2k}^H\mathbf{W}\mathbf{h}_1s+\mathbf{g}_{2k}^H\mathbf{W}\mathbf{n}_{\R}+\mathbf{g}_{2k}^H\mathbf{v}+n_{\E,2k},
\end{eqnarray}
where $n_{\E,1k}$ and $n_{\E,2k}$ are additive noise terms with zero mean and a variance of $\sigma_{\E,k}^2$.
In practice, $\eve$s typically prefer to monitor and decode the real-time data from the legitimate UEs. Therefore, it is reasonable to assume that the $\eve$s are configured on an instantaneous basis and that they rely on selection diversity combining for the sake of simplicity.
On this basis, the mutual information leakage to each $\E_k$ can therefore be expressed as
\begin{align}\label{sec2_7}
&\texttt{C}_{\E,k}(\sigma_{\Ss},\mathbf{W},\pmb{\Psi})=\frac{1}{2}\max\left\{\log_2\left(1+\frac{\sigma_{\Ss}^2|\mathbf{g}_{1k}|^2}{\sigma_{\E,k}^2}\right),\right.\nonumber\\
&\qquad~\left.\log_2\left(1+\frac{\sigma_{\Ss}^2|\mathbf{g}_{2k}^H\mathbf{W}\mathbf{h}_1|^2}{\sigma_{\R}^2\lVert\mathbf{g}_{2k}^H\mathbf{W}\rVert^2+\mathbf{g}_{2k}^H\pmb{\Psi}\mathbf{g}_{2k}+\sigma_{\E,k}^2}\right)\right\},
\end{align}
where the coefficient $\frac{1}{2}$ is due to the fact that the relay-assisted transmission requires a pair of orthogonal time slots in half-duplex mode.

In contrast to prior contributions \cite{Relay_Security_AFDF,Relay_Security_DF,Relay_Beamforming,6823730,6655008,6956810,Robust_Secure_Relaying,Robust_AF_AN}, which target maximizing the secrecy capacity, we aim for achieving the best transmission reliability, while ensuring that the information is safe to a certain extent. Specifically, assuming that in a practical communication system $\Ss$ is operating at a fixed data rate $R_d$, which is lower than its maximum achievable secrecy rate, we jointly optimize $\sigma_{\Ss}$, $\mathbf{W}$ and $\pmb{\Psi}$, subject to the power constraints, in order to maximize the received SINR at $\D$, while satisfying a set of \emph{secrecy constraints}. Mathematically, this problem can be formulated as
\begin{subequations}\label{sec2_8}
	\begin{align}
		\label{sec2_81}
		\max_{\sigma_{\Ss},\mathbf{W},\pmb{\Psi}}&\quad \frac{\sigma_{\Ss}^2|\mathbf{h}_2^H\mathbf{W}\mathbf{h}_1|^2}{\sigma_{\R}^2\lVert\mathbf{h}_2^H\mathbf{W}\rVert^2+\mathbf{h}_2^H\pmb{\Psi}\mathbf{h}_2+\sigma_{\D}^2}\\
		\label{sec2_82}
		\st&\quad \texttt{C}_{\E,k}(\sigma_{\Ss},\mathbf{W},\pmb{\Psi})\leq\kappa R_d,~k\in\mathcal{K}\\
		\label{sec2_83}
		&\quad \sigma_{\Ss}^2\lVert\mathbf{W}\mathbf{h}_1\rVert^2+\sigma_{\R}^2\lVert\mathbf{W}\rVert_F^2+\tr(\pmb{\Psi})\leq P_{\R}\\
		&\quad \sigma_{\Ss}^2\leq P_{\Ss},~\pmb{\Psi}\succeq\mathbf{0},
	\end{align}
\end{subequations}
where \eqref{sec2_82} is the so-called \emph{secrecy constraints}, which is defined from the perspective of information theoretical security \cite{shannon,hero}. Specifically, secrecy can be achieved by constraining the mutual information leakage below the data rate of the legitimate UEs. In the above formulation, the concept of a penalty $0<\kappa\leq1$ is introduced for controlling the level of security required.
Solving \eqref{sec2_8} requires perfect knowledge of all the ECSI $\left\{g_{1k},\mathbf{g}_{2k}\right\}$. However, due to the lack of explicit cooperation between the legitimate UEs and $\eve$s, only imperfect estimates of the ECSI may be available at the legitimate UEs. It is therefore important to take into consideration the uncertainties in the ECSI and to enforce the \emph{secrecy constraints} for a wide range of channel realizations. To this end, the concept of robust secrecy constraints will be introduced in the next subsection.

\subsection{Robust Secrecy Constraint}
Like most of the prior contributions in the robust transceiver design literature \cite{xing,Imperfect_CSI,Robust_Rong,Yang_TVT,Robust_AF_AN,Yang_ICC}, we model the unknown ECSI by introducing error terms $\Delta g_{1k}$ and $\Delta \mathbf{g}_{2k}$, yielding:
\begin{align}
\label{sec2_9}
g_{1k}=\hat{g}_{1k}+\Delta g_{1k},~\mathbf{g}_{2k}=\hat{\mathbf{g}}_{2k}+\Delta\mathbf{g}_{2k},
\end{align}
where $\hat{g}_{1k}$ and $\hat{\mathbf{g}}_{2k}$ denote the imperfect ECSI estimates, while again, $\Delta g_{1k}$ and $\Delta\mathbf{g}_{2k}$ capture the corresponding \emph{uncertainties}. Hereby we assume that the ECSI errors lie in some predefined bounded sets, yielding:
\begin{align}\label{sec2_10}
\mathcal{G}_{1k}&\triangleq\left\{\Delta g_{1k}:|\Delta g_{1k}|^2\leq\varepsilon_{1k}\right\}\\
\label{sec2_12}
\mathcal{G}_{2k}&\triangleq\left\{\Delta\mathbf{g}_{2k}:\lVert\Delta\mathbf{g}_{2k}\rVert^2\leq\varepsilon_{2k}\right\},
\end{align}
where $\varepsilon_{ik},~i\in\{1,2\}$ denotes the radius of the uncertainty region. The norm-bounded ECSI error model motivates the so-called \emph{worst-case} robust design approach, which seeks to maximize the received SINR, while satisfying the \emph{secrecy constraints} for any possible realization of the ECSI errors within the predefined regions. The ``robust'' version of \eqref{sec2_8} can thus be formulated as
\begin{subequations}\label{sec2_11}
	\begin{align}
		\max_{\sigma_{\Ss},\mathbf{W},\pmb{\Psi}}&\quad \SINR_{\D}\\
		\label{sec2_112}
		\st&\quad \texttt{C}_{\E,k}(\sigma_{\Ss},\mathbf{W},\pmb{\Psi};\Delta g_{1k},\Delta\mathbf{g}_{2k})\leq\kappa R_d,\nonumber\\
		&\qquad\qquad\forall \Delta g_{1k}\in\mathcal{G}_{1k},\Delta\mathbf{g}_{2k}\in\mathcal{G}_{2k},k\in\mathcal{K}\\
		\label{sec2_113}
		&\quad \sigma_{\Ss}^2\lVert\mathbf{W}\mathbf{h}_1\rVert^2+\sigma_{\R}^2\lVert\mathbf{W}\rVert_F^2+\tr(\pmb{\Psi})\leq P_{\R}\\
		&\quad \sigma_{\Ss}^2\leq P_{\Ss},~\pmb{\Psi}\succeq\mathbf{0}.
	\end{align}
\end{subequations}
One can observe that \eqref{sec2_11} is not jointly convex in $(\sigma_{\Ss},\mathbf{W},\pmb{\Psi})$ and hence in general it is difficult to solve. Additionally, the robust secrecy constraint \eqref{sec2_112} renders \eqref{sec2_11} \emph{semi-infinite}, which is mathematically challenging. To circumvent these problems, both a globally optimal and a low-complexity sub-optimal solutions will be proposed in the following sections.
%Before proceeding, we make the following remark:
%\begin{remark}
%	The considered system and channel models can properly resemble the case where $\eve$s are active entities, co-existing with the legitimate UEs within the same LTE network. A notable example is the D2D communications in the context of LTE and its evolution towards 5G \cite{D2D}. Specifically, 3GPP Release 12 defines two new functionalities, namely, \emph{device discovery} and \emph{device communication} \cite{D2D,Overview_D2D}. That is, before two UEs establish a D2D link, they can identify other UEs (potential $\eve$s) in their proximity by hearing the beacon signals, followed by an estimate of the ECSI. \hfill $\square$
%\end{remark}

\section{Globally Optimal Solution to the Robust Secrecy Problem}\label{sec3}
In this section, we develop a globally optimal solution to the robust secrecy problem \eqref{sec2_11}. Since it is challenging to jointly optimize the tuple of $\left(\sigma_{\Ss},\mathbf{W},\pmb{\Psi}\right)$ due to the non-convex nature of the problem, let us consider a sub-problem of \eqref{sec2_11}, where the aim is to solve for the optimal pair $\left(\mathbf{W},\pmb{\Psi}\right)$, while fixing the value of $\sigma_{\Ss}$. Substituting the expression of $\texttt{C}_{\E,k}$ \eqref{sec2_7} into \eqref{sec2_112} and neglecting the terms independent of $(\mathbf{W},\pmb{\Psi})$, we arrive at the sub-problem \eqref{sec3_1}, shown at the top of the next page,
\begin{figure*}[t]
	\begin{subequations}\label{sec3_1}
		\begin{align}
		\tau(\sigma_{\Ss})\triangleq\max_{\mathbf{W},\pmb{\Psi}\succeq\mathbf{0}}&\quad
		 \frac{\sigma_{\Ss}^2|\mathbf{h}_2^H\mathbf{W}\mathbf{h}_1|^2}{\sigma_{\R}^2\lVert\mathbf{h}_2^H\mathbf{W}\rVert^2+\mathbf{h}_2^H\pmb{\Psi}\mathbf{h}_2+\sigma_{\D}^2}\\
		\label{sec3_12}
		\st&\quad \sigma_{\Ss}^2\lVert\mathbf{W}\mathbf{h}_1\rVert^2+\sigma_{\R}^2\lVert\mathbf{W}\rVert_F^2+\tr(\pmb{\Psi})\leq P_{\R}\\
		\label{sec3_13}
		 &\quad\log_2\left(1+\frac{\sigma_{\Ss}^2|\mathbf{g}_{2k}^H\mathbf{W}\mathbf{h}_1|^2}{\sigma_{\R}^2\lVert\mathbf{g}_{2k}^H\mathbf{W}\rVert^2+\mathbf{g}_{2k}^H\pmb{\Psi}\mathbf{g}_{2k}+\sigma_{\E,k}^2}\right)\leq\kappa R_d,~\forall\Delta\mathbf{g}_{2k}\in\mathcal{G}_{2k},k\in\mathcal{K}.
		\end{align}
	\end{subequations}
	\hrule
\end{figure*}
where $\tau(\sigma_{\Ss})$ denotes the objective value, which depends on $\sigma_{\Ss}$. With the aid of \eqref{sec3_1}, the original problem \eqref{sec2_11} can equivalently be expressed as
\begin{align}\label{sec3_2}
	\max_{\sigma_{\Ss}}~~\tau(\sigma_{\Ss})\quad\st~~0\leq\sigma_{\Ss}\leq\bar{\sigma}_{\Ss},
\end{align}
where $\sigma_{\Ss}$ is lower bounded by zero, while the upper bound $\bar{\sigma}_{\Ss}$ can be found by solving the following problem:
\begin{align}
&\bar{\sigma}_{\Ss}=\arg\max_{\sigma_{\Ss}\leq P_{\Ss}}\quad\sigma_{\Ss}\nonumber\\
&\quad\st\max_{\Delta g_{1k}\in\mathcal{G}_{1k}}\log_2\left(1+\frac{\sigma_{\Ss}^2|g_{1k}|^2}{\sigma_{\E,k}^2}\right)\leq\kappa R_d,k\in\K.
\end{align}
The resultant solution is given by
\begin{equation}\label{sec3_9}
	 \bar{\sigma}_{\Ss}=\min\left\{\sqrt{P_{\Ss}},\min_{k\in\K}\left\{\frac{\gamma\sigma_{\E,k}^2}{\left||\hat{g}_{1k}|+\sqrt{\varepsilon_{1k}}\right|^2}\right\}\right\},
\end{equation}
where $\gamma\triangleq 2^{2\kappa R_d}-1$. The reformulated problem \eqref{sec3_2} leads to a simpler single-variable optimization problem defined over the interval $[0,~\bar{\sigma}_{\Ss}]$. Assuming that $\tau(\sigma_{\Ss})$ can be evaluated at any feasible $\sigma_{\Ss}$, a one-dimensional exhaustive search procedure can be invoked for finding the global optimum of \eqref{sec2_11}. Let us now focus our attention on computing $\tau(\sigma_{\Ss})$ for a given feasible $\sigma_{\Ss}$, which however requires solving the non-convex sub-problem \eqref{sec3_1}. The solution to \eqref{sec3_1} will be addressed in the following.

We begin by tackling the challenge posed by the infinite number of constraints in \eqref{sec3_13}. After some further manipulations, we can rewrite \eqref{sec3_13} as
\begin{multline}\label{sec3_3}
\Delta\mathbf{g}_{2k}^H\pmb{\Theta}(\mathbf{W},\pmb{\Psi})\Delta\mathbf{g}_{2k}+2\rea\left\{\hat{\mathbf{g}}_{2k}^H\pmb{\Theta}(\mathbf{W},\pmb{\Psi})\Delta\mathbf{g}_{2k}\right\}\\
+\hat{\mathbf{g}}_{2k}^H\pmb{\Theta}(\mathbf{W},\pmb{\Psi})\hat{\mathbf{g}}_{2k}-\gamma\sigma_{\E,k}^2\leq 0,~\forall \Delta\mathbf{g}_{2k}\in\mathcal{G}_{2k},
\end{multline}
where we have $\pmb{\Theta}(\mathbf{W},\pmb{\Psi})\triangleq\mathbf{W}\left(\sigma_{\Ss}^2\mathbf{h}_1\mathbf{h}_1^H-\gamma\sigma_{\R}^2\mathbf{I}_{N_{\R}}\right)\mathbf{W}^H-\gamma\pmb{\Psi}$.
As a popular technique of tackling the infiniteness, we invoke the so-called $\mathcal{S}$-Procedure \cite{Convex_Optimization} for equivalently recasting \eqref{sec3_3} as
%\begin{lemma}[$\mathcal{S}$-Procedure \cite{Convex_Optimization}]\label{lemma1}
%Consider $\mathbf{A}=\mathbf{A}^H$, $\mathbf{D}=\mathbf{D}^H\in\mathbb{C}^{N\times N}$, $\mathbf{b}\in\mathbb{C}^{N\times1}$, and assume the interior condition holds, i.e., there exists an $\bar{\mathbf{x}}\in\mathbb{C}^{N\times1}$ such that $\bar{\mathbf{x}}^H\mathbf{D}\bar{\mathbf{x}}<e$. Then the following two conditions are equivalent:
%\begin{enumerate}
%    \item $\mathbf{x}^H\mathbf{A}\mathbf{x}+2\rea(\mathbf{b}^H\mathbf{x})+c\leq 0,~\forall\mathbf{x}^H\mathbf{D}\mathbf{x}\leq e$;
%    \item There exists $\theta\geq0$ such that
%    \begin{equation}\label{e30}
%    \left[\begin{array}{cc}
%        \theta\mathbf{D}-\mathbf{A} & -\mathbf{b}\\
%        -\mathbf{b}^H & -c-e\theta
%        \end{array}\right]\succeq\mathbf{0}.
%    \end{equation}
%\end{enumerate}
%\end{lemma}
\begin{align}\label{sec3_4}
\mathbf{M}_k(\mathbf{W},\pmb{\Psi},\rho_k)&\triangleq\pmb{\Lambda}_k(\rho_k)-\mathbf{P}_k^H\pmb{\Theta}(\mathbf{W},\pmb{\Psi})\mathbf{P}_k\succeq\mathbf{0},
\end{align}
where we have $\pmb{\Lambda}_k(\rho_k)=\blkdiag(\rho_k\mathbf{I}_{N_{\R}},\gamma\sigma_{\E,k}^2-\varepsilon_{2k}\rho_k)$ and $\mathbf{P}_k=[\mathbf{I}_{N_{\R}},\hat{\mathbf{g}}_{2k}]$.
Replacing \eqref{sec3_12} by \eqref{sec3_4}, the sub-problem \eqref{sec3_1} can now be rewritten in a \emph{finite} form as
\begin{subequations}\label{sec3_5}
\begin{align}
	\label{sec3_51}
	\max_{\mathbf{W},\pmb{\Psi}}&\quad \frac{\sigma_{\Ss}^2|\mathbf{h}_2^H\mathbf{W}\mathbf{h}_1|^2}{\sigma_{\R}^2\lVert\mathbf{h}_2^H\mathbf{W}\rVert^2+\mathbf{h}_2^H\pmb{\Psi}\mathbf{h}_2+\sigma_{\D}^2}\\
	\label{sec3_52}
	\st&\quad \sigma_{\Ss}^2\lVert\mathbf{W}\mathbf{h}_1\rVert^2+\sigma_{\R}^2\lVert\mathbf{W}\rVert_F^2+\tr(\pmb{\Psi})\leq P_{\R}\\
	\label{sec3_53}
	&\quad\pmb{\Lambda}_k(\rho_k)-\mathbf{P}_k^H\pmb{\Theta}(\mathbf{W},\pmb{\Psi})\mathbf{P}_k\succeq\mathbf{0},~k\in\mathcal{K}\\
	&\quad\pmb{\Psi}\succeq\mathbf{0}.
\end{align}
\end{subequations}
The above formulation remains non-convex and to proceed, we have to transform it into an appropriate form, where the SDR is applicable. Let us define $\mathbf{w}=\vect(\mathbf{W})$ and $\mathbf{X}=\mathbf{w}\mathbf{w}^H$. Interestingly, after some matrix manipulations, which have been relegated to Appendix~\ref{appendix1}, we are now able to rewrite \eqref{sec3_5} in a form, which only involves linear terms in $\mathbf{X}$ and $\pmb{\Psi}$. The results are summarized in the following proposition:
\begin{proposition}\label{proposition1}
Define
\begin{align}
\mathbf{Q}_0&=\sigma_{\Ss}^2(\mathbf{h}_1^*\mathbf{h}_1^T)\otimes(\mathbf{h}_2\mathbf{h}_2^H)\\
\mathbf{Q}_1&=\sigma_{\R}^2\mathbf{I}_{N_{\R}}\otimes(\mathbf{h}_2\mathbf{h}_2^H)\\
\mathbf{Q}_2&=\sigma_{\Ss}^2(\mathbf{h}_1^*\mathbf{h}_1^T)\otimes\mathbf{I}_{N_{\R}}+\sigma_{\R}^2\mathbf{I}_{N_{\R}^2}\\
\mathbf{Q}_3(\mathbf{X},\pmb{\Psi})&=\sigma_{\Ss}^2\mathbf{H}_1\mathbf{X}\mathbf{H}_1^H-\gamma\sigma_{\R}^2\sum_{l=1}^{N_{\R}}\mathbf{E}_l\mathbf{X}\mathbf{E}_l^H-\gamma\pmb{\Psi},
\end{align}
where $\mathbf{Q}_3(\cdot)$ is a linear mapping of $\mathbf{X}$ and $\pmb{\Psi}$ with $\mathbf{H}_1=\mathbf{h}_1^T\otimes\mathbf{I}_{N_{\R}}$ and $\mathbf{E}_l=\left[\mathbf{0}_{N_{\R}\times(l-1)N_{\R}},\mathbf{I}_{N_{\R}},\mathbf{0}_{N_{\R}\times(N_{\R}-l)N_{\R}}\right]$.
Then, problem \eqref{sec3_5} can equivalently be rewritten as follows:
\begin{subequations}\label{sec3_6}
	\begin{align}
		\label{sec3_61}
		\max_{\mathbf{X},\pmb{\Psi},\pmb{\rho}}&\quad \frac{\tr(\mathbf{Q}_0\mathbf{X})}{\tr(\mathbf{Q}_1\mathbf{X})+\tr(\mathbf{h}_2\mathbf{h}_2^H\pmb{\Psi})+\sigma_{\D}^2}\\
		\label{sec3_62}
		\st&\quad \tr(\mathbf{Q}_2\mathbf{X})+\tr(\pmb{\Psi})\leq P_{\R}\\
		\label{sec3_63}
		&\quad\pmb{\Lambda}_k(\rho_k)-\mathbf{P}_k^H\mathbf{Q}_3(\mathbf{X},\pmb{\Psi})\mathbf{P}_k\succeq\mathbf{0},~k\in\mathcal{K}\\
		\label{sec3_64}
		&\quad\mathbf{X}\succeq\mathbf{0},~\pmb{\Psi}\succeq\mathbf{0},~\rank(\mathbf{X})=1.
	\end{align}
\end{subequations}
\end{proposition}
Upon neglecting the non-convex rank-one constraint in \eqref{sec3_64}, \eqref{sec3_6} is relaxed to a so-called fractional SDP, which can further be recast into a standard SDP via the Charnes-Cooper transformation \cite{Charnes_Cooper}. Specifically, by introducing an auxiliary variable $s>0$, and defining $\overline{\mathbf{X}}=s\mathbf{X}$, $\overline{\pmb{\Psi}}=s\pmb{\Psi}$ and $\overline{\pmb{\rho}}=s\pmb{\rho}$, \eqref{sec3_6} is conveniently transformed into
\begin{subequations}
\label{sec3_7}
\begin{align}
\max_{\overline{\mathbf{X}},\overline{\pmb{\Psi}},\overline{\pmb{\rho}},s>0}&\quad\tr(\mathbf{Q}_0\overline{\mathbf{X}})\\
\st&\quad\tr(\mathbf{Q}_1\overline{\mathbf{X}})+\tr(\mathbf{h}_2\mathbf{h}_2^H\overline{\pmb{\Psi}})+s\sigma_{\D}^2\leq1\\
&\quad \tr(\mathbf{Q}_2\overline{\mathbf{X}})+\tr(\overline{\pmb{\Psi}})\leq sP_{\R}\\
\label{sec3_73}
&\quad\pmb{\Lambda}_k(\overline{\rho}_k)-\mathbf{P}_k^H\mathbf{Q}_3(\overline{\mathbf{X}},\overline{\pmb{\Psi}})\mathbf{P}_k\succeq\mathbf{0},~k\in\mathcal{K}\\
&\quad\overline{\mathbf{X}}\succeq\mathbf{0},~\overline{\pmb{\Psi}}\succeq\mathbf{0}.
\end{align}
\end{subequations}
Interestingly, \eqref{sec3_7} now becomes a convex SDP, which can be efficiently solvable by generic optimization tools such as \texttt{SeDuMi} \cite{SeDuMi} and \texttt{MOSEK} \cite{mosek} by relying on interior-point methods \cite{Interior_Point}. We remark that \eqref{sec3_7} and the rank-relaxed version of \eqref{sec3_6} are equivalent in the sense that the optimal solution $\mathbf{X}^*$ to \eqref{sec3_6} after rank-one relaxation can be retrieved from the optimal solution $(\overline{\mathbf{X}}^*,s^*)$ to \eqref{sec3_7}, i.e., $\mathbf{X}^*=\overline{\mathbf{X}}^*/s^*$, and the resultant objective values of the two problems are equivalent (see \cite{Robust_AF_AN} for a detailed proof of this equivalence).

After obtaining the rank-relaxed solution $\mathbf{X}^*$, a natural question arises as to how good a solution is  $\mathbf{X}^*$, i.e., does it satisfy the rank-one optimality condition of \eqref{sec3_6}? Answering this question directly from the formulation of \eqref{sec3_7} is still an open problem in the literature. To overcome this difficulty, we follow an approach similar to \cite{Robust_AF_AN}. Specifically, denoting the objective value of \eqref{sec3_7} by $\tau_{\text{relax}}^*(\sigma_{\Ss})$, we consider the following power minimization problem:
\begin{subequations}\label{sec3_8}
\begin{align}
\min_{\mathbf{X},\pmb{\Psi},\pmb{\rho}}&\quad\tr(\mathbf{Q}_2\mathbf{X})\\
\st&\quad\frac{\tr(\mathbf{Q}_0\mathbf{X})}{\tr(\mathbf{Q}_1\mathbf{X})+\tr(\mathbf{h}_2\mathbf{h}_2^H\pmb{\Psi})+\sigma_{\D}^2}\geq\tau_{\text{relax}}^*(\sigma_{\Ss})\\
&\quad \tr(\mathbf{Q}_2\mathbf{X})+\tr(\pmb{\Psi})\leq P_{\R}\\
\label{sec3_83}
&\quad\pmb{\Lambda}_k(\rho_k)-\mathbf{P}_k^H\mathbf{Q}_3(\mathbf{X},\pmb{\Psi})\mathbf{P}_k\succeq\mathbf{0},~k\in\mathcal{K}\\
&\quad\mathbf{X}\succeq\mathbf{0},~\pmb{\Psi}\succeq\mathbf{0}.
\end{align}
\end{subequations}
Observe that \eqref{sec3_8} is also a standard SDP and therefore it is readily solvable by existing optimization tools. Furthermore, its specific structure allows us to obtain the following useful result, based on which we will be able to retrieve an optimal rank-one solution of \eqref{sec3_6}.
\begin{proposition}\label{proposition2}
Let us denote the optimal solution of \eqref{sec3_8} by $(\mathbf{X}^o,\pmb{\Psi}^o,\pmb{\rho}^o)$. Assuming suitable constraint qualification of \eqref{sec3_8}, $(\mathbf{X}^o,\pmb{\Psi}^o,\pmb{\rho}^o)$ is also an optimal solution of \eqref{sec3_6}, i.e., $\mathbf{X}^o$ must be of rank one.
\end{proposition}

The proof follows arguments similar to those in \cite[Appendices B--C]{Robust_AF_AN}, which is therefore omitted for brevity. The key ingredient of the proof is to show that $\mathbf{X}^o$ is of rank one, which can be achieved by examining the Karush-Kuhn-Tucker (KKT) conditions for \eqref{sec3_8}.

In summary, obtaining an optimal solution of \eqref{sec3_2} now consists of two steps: \emph{1)} solve the rank-relaxed SDP \eqref{sec3_7} and obtain the largest $\tau_{\text{relax}}^*(\sigma_{\Ss})$ by exhaustive search over $\sigma_{\Ss}$; \emph{2)} solve the power minimization problem \eqref{sec3_8} based on the obtained $\tau_{\text{relax}}^*(\sigma_{\Ss})$. Since the rank-one optimality condition of $\mathbf{X}^o$ is guaranteed, the optimal AF matrix $\mathbf{W}^o$ can be retrieved by rank-one decomposition of $\mathbf{X}^o$, i.e., $\mathbf{X}^o=\mathbf{x}^o(\mathbf{x}^o)^H$, followed by converting $\mathbf{x}^o$ to $\mathbf{W}^o$ via vector-matrix reshaping.
%Setting a step size $\Delta P$, the iterative algorithm relying on the one-dimensional exhaustive search is summarized as Algorithm~\ref{alg1}.
%\begin{algorithm}[h]
%	\KwIn{$\Delta$, $I=\left\lfloor\frac{\bar{\sigma}_{\Ss}}{\Delta}\right\rfloor$, $\mathcal{S}=\emptyset$}
%	\For{$i=1,\cdots,I$}{
%		$\sigma_{\Ss}^*=i\cdot\Delta$\;
%		Solve for $(\mathbf{X}^*,\pmb{\Psi}^*)$\;
%		\eIf{$\rank(\mathbf{X}^*)>1$}{
%			Solve for $(\mathbf{X}^*,\pmb{\Psi}^*)$\;
%			Obtain $\mathbf{W}^*=\mathtt{MAT}(\mathtt{devec}(\mathbf{X}^*))$\;
%		}{
%			Obtain $\mathbf{W}^*=\mathtt{MAT}(\mathtt{devec}(\mathbf{X}^*))$\;
%		}
%		$\mathcal{S}\leftarrow (\sigma_{\Ss}^*,\mathbf{W}^*),\pmb{\Psi}^*$\;
%	}
%	 \KwOut{$(P_{\Ss}^{\text{opt}},\mathbf{W}^{\text{opt}})=\arg\max\limits_{(P_{\Ss}^*,\mathbf{W}^*)\in\mathcal{S}}~\mathtt{SINR}_{\D}(P_{\Ss}^*,\mathbf{W}^*)$}
%	 \label{algo1}
%\end{algorithm}

We emphasize that solving \eqref{sec3_2} requires performing an exhaustive search for $\sigma_{\Ss}$ over $[0,~\bar{\sigma}_{\Ss}]$. In each step, we have to solve the SDP \eqref{sec3_7}, which involves on the order of $\mathcal{O}(N_{\R}^4+N_{\R}^2+1)$ optimization variables and $K$ semidefinite cone constraints of dimension $(N_{\R}+1)^2$. Hence, \eqref{sec3_7} can be solved with a \emph{worst-case} complexity, which is on the order of $\mathcal{O}\left(K(N_{\R}^4+N_{\R}^2+1)^2(N_{\R}+1)^2\right)$ \cite{SDP}. The associated computational cost escalates as the number of relay antennas or $\eve$s increases, which may become computationally prohibitive in practical problems. Motivated by this issue, a low-complexity sub-optimal algorithm will be conceived in the next section.

\section{A Penalized DC Algorithmic Framework}\label{sec4}
Achieving the global optimum presented in Section~\ref{sec3} may only become computationally affordable when the problem size is small; nevertheless, it remains useful in general as a baseline for benchmarking other algorithms. As an alternative, we propose here a low-complexity algorithm by resorting to a new penalized DC algorithmic framework. We first expose this framework, which can be considered as an extended variation of the conventional DC framework \cite{DC_Overview}.
Then as a second contribution, since the results of convergence analysis in the conventional DC are not directly applicable to this extended framework, we explicitly state and prove the convergence properties of the new penalized DC algorithm.
%Finally, to further reduce the computational complexity, an \emph{inexact} version of the latter is proposed.

\subsection{Penalized DC Algorithmic Framework}
We first introduce a few definitions prior to the algorithm design:
\begin{definition}[Positive Semi-Definite (PSD)-Convex Mapping]
A matrix-valued mapping $\pmb{\mathcal{F}}(\cdot):\mathbb{C}^n\rightarrow\mathbb{H}^p$ is called PSD-convex on a convex subset $\Omega\subseteq\mathbb{C}^n$, if for all $\mathbf{x},\mathbf{y}\in\Omega$ and $\theta$ with $0\leq\theta\leq1$, we have
\begin{equation}\label{sec4_1}
\pmb{\mathcal{F}}\left(\theta\mathbf{x}+(1-\theta)\mathbf{y}\right)\preceq \theta\pmb{\mathcal{F}}(\mathbf{x})+(1-\theta)\pmb{\mathcal{F}}(\mathbf{y}).
\end{equation}
\end{definition}
\begin{definition}[Directional Derivative of Matrix-Valued Mapping]
The directional derivative of a matrix-valued mapping $\pmb{\mathcal{F}}$ at $\mathbf{x}$ is a linear mapping $\pmb{\mathcal{DF}}: \mathbb{C}^{n}\rightarrow\mathbb{H}^p$, which is defined by
\begin{equation}\label{sec4_2}
\pmb{\mathcal{DF}}\mathbf{h}=\sum_{i=1}^nh_i\frac{\partial\pmb{\mathcal{F}}}{\partial x_i}(\mathbf{x}),~\forall \mathbf{h}\in\mathbb{C}^n.
\end{equation}
\end{definition}
\begin{definition}[PSD-DC Program]
	A PSD-DC program assumes the form of
	\begin{subequations}\label{sec4_3}
		\begin{align}
			\label{sec4_31}
			\min_{\mathbf{x}\in\Omega}&\quad \varphi(\x)\triangleq f_0(\mathbf{x})-g_0(\mathbf{x})\\
			\label{sec4_32}
			\st&\quad \pmb{\mathcal{F}}_i(\mathbf{x})-\pmb{\mathcal{G}}_i(\mathbf{x})\preceq\mathbf{0},~i\in\mathcal{I}\triangleq\{1,2,\cdots,I\}
			\end{align}
	\end{subequations}
	where $\Omega\subseteq\mathbb{C}^n$ is a non-empty, closed convex set, $f_0(\cdot)$, $g_0(\cdot): \mathbb{C}^n\rightarrow\mathbb{R}$ are continuously differentiable convex functions on $\Omega$, and $\pmb{\mathcal{F}}_i(\cdot)$, $\pmb{\mathcal{G}}_i(\cdot), \mathbb{C}^n\rightarrow\mathbb{H}^{p_i}$ are continuously differentiable PSD-convex mappings on $\Omega$.
\end{definition}

We know that a PSD-convex mapping provides a generalization of a convex function, since any convex function with $f(\cdot):\mathbb{C}^n\rightarrow\mathbb{R}$ is PSD-convex in conjunction with $p=1$. Similarly, a PSD-DC program generalizes a conventional DC program, where an inequality constraint, e.g., $f_i(\mathbf{x})-g_i(\mathbf{x})\leq 0$ is now extended to a \emph{generalized inequality} $\preceq$ on a PSD cone. If $g_0(\cdot)$ is nonlinear and one of the $\G_i(\cdot)$ is a nonlinear mapping, then \eqref{sec4_3} becomes a general non-convex nonlinear SDP.

To solve \eqref{sec4_3}, similar to the conventional DC framework \cite{CCP} which involves only \emph{scalar-valued functions}, an iterative algorithm can be developed, where the main idea is to find a local linear approximation of the non-convex parts of \eqref{sec4_3} around the solution obtained in the previous iteration.
In this way, the original non-convex problem can then be iteratively solved by a sequence of ``convexified'' sub-problems.
Assuming that $\mathbf{x}^{(n)}$ is a solution obtained at the $n$\textsuperscript{th} iteration,
the linearized sub-problem is then given by
\begin{subequations}\label{sec4_4}
    \begin{align}
        \label{sec4_41}
        \min_{\mathbf{x}\in\Omega}&\quad f_0(\mathbf{x})-g_0(\mathbf{x}^{(n)})-\nabla g_0^T(\mathbf{x}^{(n)})(\mathbf{x}-\mathbf{x}^{(n)})\\
        \label{sec4_42}
        \st&\quad\F_i(\x)-\G_i(\x^{(n)})-\pmb{\mathcal{DG}}_i(\x^{(n)})(\x-\x^{(n)})\preceq\mathbf{0},~i\in\mathcal{I}.
    \end{align}
\end{subequations}
In order to generate a sequence of feasible solutions $\{\x^{(n)}\}_{n=0}^{\infty}$, a feasible initialization $\x^{(0)}$ is required by the iterative algorithm\footnote{This is due to the fact that the first-order Taylor series expansion of the concave function $-\G_i(\cdot)$ is its upper bound, such that we have $\F_i(\x)-\G_i(\x)\preceq\F_i(\x)-\G_i(\x^{(n)})-\pmb{\mathcal{DG}}_i(\mathbf{x}^{(n)})(\mathbf{x}-\mathbf{x}^{(n)})\preceq\mathbf{0}$ for all $\mathbf{x}\in\Omega$.}.
Otherwise, if the algorithm starts with an infeasible point, it can lead to further infeasibility problems during the successive iterations.
However, finding a ``good'' feasible initialization for a non-convex problem in principle is not a simple task,
as argued further in Section~\ref{sec5}.

Motivated by the above considerations, we propose an alternative approach relying on the concept of penalized DC algorithm, which now \textbf{\emph{eliminates the requirement of a non-trivial initialization}}.
Instead of solving \eqref{sec4_4}, hereby we introduce a set of matrix auxiliary variables $\{\mathbf{S}_i\}_{i=1}^I$ and penalize \eqref{sec4_41} with a linear regularization term, i.e.,
\begin{subequations}\label{sec4_5}
	\begin{align}
		\label{sec4_51}
		\min_{\mathbf{x}\in\Omega,\mathbf{S}}&\quad \hat{\varphi}^{(n)}(\x,\mathbf{S};\xn)\nonumber\\
		&\quad\quad\triangleq f_0(\mathbf{x})-g_0(\mathbf{x}^{(n)})\nonumber\\
		&\qquad\quad~-\nabla g_0^T(\mathbf{x}^{(n)})(\mathbf{x}-\mathbf{x}^{(n)})+\tau^{(n)}\sum_{i=1}^I\tr(\mathbf{S}_i)\\
		\label{sec4_52}
		\st&\quad \pmb{\mathcal{F}}_i(\mathbf{x})-\pmb{\mathcal{G}}_i(\mathbf{x}^{(n)})-\pmb{\mathcal{DG}}_i(\mathbf{x}^{(n)})(\mathbf{x}-\mathbf{x}^{(n)})\preceq\mathbf{S}_i\\
		\label{sec4_53}
		&\quad\mathbf{S}_i\succeq\mathbf{0},~i\in\mathcal{I}
	\end{align}
\end{subequations}
where $\tau^{(n)}\geq0$ denotes the weight associated with the penalty term at the $n$\textsuperscript{th} iteration and $\mathbf{S}\triangleq(\mathbf{S}_1,\cdots,\mathbf{S}_I)$. Auxiliary matrix $\mathbf{S}_i\in\mathbb{H}^{p_i}$ can be viewed as an abstract measure of the extent to which the $i$\textsuperscript{th} constraint in \eqref{sec4_42} is violated. The feasibility indicator $\tr(\mathbf{S}_i)=0$ reveals that the $i$\textsuperscript{th} constraint is satisfied while $\tr(\mathbf{S}_i)>0$ indicates the opposite. Therefore, a feasible solution $\mathbf{x}\in\Omega$ is found if $\sum_{i=1}^I\tr(\mathbf{S}_i)=0$.

\begin{algorithm}[t]
	\caption{Penalized DC Algorithm}
	\label{alg1}
	\begin{algorithmic}\raggedright
		\Require An initial point $\mathbf{x}^{(0)}\in\Omega$, $\tau^{(0)}>0$, $\delta_1>0$ and $\delta_2>0$. Set $n=0$.
		\Repeat
		\State 1. \emph{Convexify}: Compute the first-order approximates
		\begin{align*}
		g_0(\mathbf{x})&\approx g_0\big(\mathbf{x}^{(n)}\big)+\nabla g_0^T\big(\mathbf{x}^{(n)}\big)\big(\mathbf{x}-\mathbf{x}^{(n)}\big)\\
		\pmb{\mathcal{G}}_i\big(\mathbf{x}\big)&\approx\pmb{\mathcal{G}}_i\big(\mathbf{x}^{(n)}\big)+\pmb{\mathcal{DG}}_i\big(\mathbf{x}^{(n)}\big)\big(\mathbf{x}-\mathbf{x}^{(n)}\big)
		\end{align*}
		\State 2. \emph{Solve}: Compute $\mathbf{x}^{(n+1)}$ by solving \eqref{sec4_5}
		\State 3. \emph{Update} $\tau$: Obtain the dual variable $\pmb{\Phi}_i^{(n+1)}$ associated with \eqref{sec4_52} and set
		\begin{equation}\label{sec4_alg1}
		\tau^{(n+1)}=\left\{
		\begin{array}{ll}
		\tau^{(n)} & \text{if}~\tau^{(n)}\geq r^{(n)}\\
		\tau^{(n)}+\delta_2 & \text{if}~\tau^{(n)}<r^{(n)}
		\end{array}\right.
		\end{equation}
		where
		\begin{equation*}
		r^{(n)}\triangleq\min\left\{\lVert\mathbf{x}^{(n+1)}-\mathbf{x}^{(n)}\rVert^{-1},\lambda_{\max}\Big[\sum_{i=1}^I\pmb{\Phi}_i^{(n+1)}\Big]+\delta_1\right\}
		\end{equation*}
		\State 4. \emph{Update iteration}: $n \leftarrow n+1$
		\Until{Termination criterion is satisfied \emph{or} a maximum number of iterations is reached}
		\Ensure The optimized $\mathbf{x}^*$.
	\end{algorithmic}
\end{algorithm}

The proposed penalized DC algorithm, which iteratively solves a sequence of sub-problems \eqref{sec4_5} with a specific updating rule of $\tau$ is listed as $\algone$. The rationale for this algorithm is that it starts with a possibly infeasible point with a small penalty $\tau$. In turn, this enables a fast descent of the objective function at the beginning while the constraints can be temporarily violated, i.e, $\mathbf{S}_i\succ\mathbf{0}$. As the number of iterations increases, $\tau$ gradually increases, thereby forcing the solution to be closer to and finally lie in the feasible region.

\begin{remark}
Before leaving this subsection, we discuss a few important aspects of $\algone$.

\emph{1) Initialization:} Instead of finding an initialization in the non-convex feasible set of \eqref{sec4_3}, i.e., $\mathbf{x}^{(0)}\in\mathcal{F}\triangleq\{\x\in\Omega|\F_i(\x)-\G_i(\x)\preceq\mathbf{0},~i\in\mathcal{I}\}$, $\algone$ is now initialized with a point $\mathbf{x}^{(0)}\in\Omega$, which leads to a more computationally efficient \emph{convex} feasible search problem. For implementation, one may rely on general-purpose optimization solvers to find $\x^{(0)}$. More importantly, in many practical problems, $\x^{(0)}$ can be easily found by exploiting the specific structure of $\Omega$. This is illustrated in Section~\ref{sec5} for the robust secure relaying design problem.

\emph{2) Termination Criterion:} In practice, $\algone$ needs to be terminated within a maximum of number iterations. For instance, a reasonable termination criterion is that the difference between successive solutions becomes small, i.e., $\lVert\xnn-\xn\rVert\leq\delta$ and $\xn$ is (nearly) feasible, i.e., $\sum_{i=1}^I\tr(\mathbf{S}_i)\approx0$. If this criterion cannot be satisfied within a preset number of iterations, we claim that the algorithm fails to find a feasible solution within a limited time frame.
%though we are able to prove that any limit point of $\algone$ is a stationary point of \eqref{sec4_3}.

\emph{3) On the Updating Rule:} The updating rule of $\tau$ in \eqref{sec4_alg1} is motivated by the theory of exact penalty functions for nonlinear optimization problems \cite{Penalty_Nonlinear,Penalty_Constrained}. The theory suggests that if the penalty $\tau$ is larger than all the dual variables $\{\pmb{\Phi}_i\}$ associated with \eqref{sec4_52} (in our case, it is in the form of PSD ordering), i.e., $\tau\mathbf{I}\succeq\pmb{\Phi}_i$ for all $i$, then \eqref{sec4_5} and \eqref{sec4_4} become equivalent.
Also from the definition of $r^{(n)}$ below \eqref{sec4_alg1}, we see that the unboundedness of $\{\tau^{(n)}\}$ leads to the unboundedness of $\{\pmb{\Phi}_i^{(n)}\}$ as well as $\lVert\mathbf{x}^{(n+1)}-\mathbf{x}^{(n)}\rVert\rightarrow 0$. This key property will be exploited later in proving the convergence of Algorithm~\ref{alg1}.

\emph{4) Solving the Sub-Problem \eqref{sec4_5}:} As mentioned earlier, \eqref{sec4_5} is a general convex problem. Specifically, if $\Omega$ can be represented by a number of finite linear matrix inequalities (LMIs), and if $f_0(\cdot)$ or one of the mappings $\F_i(\cdot)$ is nonlinear in $\mathbf{x}$, then \eqref{sec4_5} becomes a \emph{nonlinear} convex SDP.
To the best of our knowledge, the availability of external solvers supporting a general \emph{nonlinear} SDP is still limited at the time of writing. Briefly, the widely-used solvers such as \texttt{SeDuMi} and \texttt{MOSEK} do not support nonlinear SDPs. However, it has been shown in the literature that many beamforming/MIMO transceiver design problems exhibit some common structures, namely: \emph{1)} $f_0(\cdot)$ is a quadratic function of $\mathbf{x}$, i.e., $f_0(\mathbf{x})=\mathbf{x}^H\mathbf{P}\mathbf{x}$ where $\mathbf{P}\succeq\mathbf{0}$; \emph{2)} $\F_i(\cdot)$ is referred to as being Schur PSD-convex, which can be expressed as $\F_i(\mathbf{x})=\mathbf{A}_i(\mathbf{x})\mathbf{Q}_i^{-1}\mathbf{A}_i^H(\mathbf{x})-\mathbf{B}_i(\mathbf{x})$, where $\mathbf{A}_i(\mathbf{x})$ and $\mathbf{B}_i(\mathbf{x})$ are linear mappings of $\mathbf{x}$. In this case, we can invoke the Schur complement rule
for transforming \eqref{sec4_5} into a standard SDP as follows:
\begin{flalign*}
	&\min_{\substack{\mathbf{x}\in\Omega,\mathbf{S},\\ t,\{\mathbf{T}_i\}}}\quad t-g_0(\mathbf{x}^{(n)})-\nabla g_0^T(\mathbf{x}^{(n)})(\mathbf{x}-\mathbf{x}^{(n)})+\tau^{(n)}\sum_{i=1}^I\tr(\mathbf{S}_i)&\\
	&\quad\st\quad
	\left[\begin{array}{cc}
	\mathbf{P} & \x\\
	\x^H & t
	\end{array}\right]\succeq\mathbf{0}&\\	
	&\quad\phantom{\st}\quad
	\mathbf{T}_i-\pmb{\mathcal{G}}_i(\mathbf{x}^{(n)})-\pmb{\mathcal{DG}}_i(\mathbf{x}^{(n)})(\mathbf{x}-\mathbf{x}^{(n)})\preceq\mathbf{S}_i&\\
	&\quad\phantom{\st}\quad
	\left[\begin{array}{cc}
			\mathbf{Q}_i & \mathbf{A}_i^H(\x)\\
			\mathbf{A}_i(\x) & \mathbf{T}_i
		\end{array}\right]\succeq\mathbf{0},~\mathbf{S}_i\succeq\mathbf{0},~i\in\mathcal{I},&
\end{flalign*}
where $t$ and $\{\mathbf{T}_i\}$ are auxiliary variables. The above problem can be efficiently solved both by \texttt{SeDuMi} and \texttt{MOSEK}.
\end{remark}

\subsection{Convergence Analysis of the Penalized DC Algorithm}
Since \eqref{sec4_5} may admit an infeasible solution to the original problem \eqref{sec4_3}, two important aspects regarding the convergence of Algorithm~\ref{alg1} have to be examined:
\emph{1)} whether the solution generated by $\algone$ is feasible for \eqref{sec4_3}?
and \emph{2)} if the convergence properties of the conventional DC algorithm still hold for the penalized DC algorithm?
In this subsection, the convergence properties of $\algone$ are analytically established under the following assumptions\footnote{We remark that A.1)--A.3) are mild assumptions from both theoretical and practical perspectives. In fact, A.1) is the so-called extended Magsanrian-Fromovitz constraint qualification, which is considered to be a rather mild condition for classic non-convex nonlinear optimization problems \cite{Penalty_Nonlinear}. In A.2), $\Omega$ is bounded due to the power constraints imposed in the design problem, while the objective function is usually a performance metric such as the SINR or the mean-square-error (MSE), which are lower-bounded by zero. Regarding A.3), consider the objective function $\varphi(\x)$ \eqref{sec4_31} which can be equivalently written as $\varphi_(\x)= (f_0(\mathbf{x})+\frac{\rho}{2}\lVert\cdot\rVert^2)-(g_0(\mathbf{x})+\frac{\rho}{2}\lVert\cdot\rVert^2)$ for any given $\rho>0$. Therefore, without loss of generality, we can always find a DC decomposition $(f_0, g_0)$  such that both $f_0$ and $g_0$ are strongly convex.}:
\begin{itemize}%\setlength{\itemindent}{+.5in}
	\item [A.1)] For any $\mathbf{x}\in\Omega$, there exists a vector $\mathbf{h}\in\texttt{cone}(\Omega-\mathbf{x})$ such that
	\begin{equation}
		\big(\pmb{\mathcal{DF}}_i(\mathbf{x})-\pmb{\mathcal{DG}}_i(\mathbf{x})\big)\mathbf{h}\prec\mathbf{0},~\forall i\in\mathcal{U}(\mathbf{x})
	\end{equation}
	where $\mathcal{U}(\mathbf{x})\triangleq\left\{i\in\mathcal{I}\big|\pmb{\mathcal{F}}_i(\mathbf{x})-\pmb{\mathcal{G}}_i(\mathbf{x})\nprec\mathbf{0}\right\}$;
	\item [A.2)] $\Omega$ is bounded and the objective function $f_0(\x)-g_0(\x)$ is bounded from below;
	\item [A.3)] Either $f_0(\cdot)$ or $g_0(\cdot)$ is strongly convex\footnote{A function $f(\cdot)$ is said to be strongly convex with parameter $\rho>0$, if the following inequality holds for all $\mathbf{x},\mathbf{y}$ in its domain: $\left(\nabla f(\mathbf{x})-\nabla f(\mathbf{y})\right)^T\left(\mathbf{x}-\mathbf{y}\right)\geq\rho\lVert\mathbf{x}-\mathbf{y}\rVert^2$, or equivalently, $ f(\mathbf{y})\geq f(\mathbf{x})+\nabla f(\mathbf{x})^T(\mathbf{y}-\mathbf{x})+\frac{\rho}{2}\lVert\mathbf{x}-\mathbf{y}\rVert^2$.}.
\end{itemize}
\begin{theorem}\label{theorem1}
	Let $\left\{\mathbf{x}^{(n)}\right\}$ be the solution sequence generated by $\algone$. Assuming that \eqref{sec4_3} is feasible and A.1)--A.3) hold for \eqref{sec4_3}. Then, one of the following scenarios applies:
	\begin{itemize}
		\item[\emph{1)}] Algorithm~\ref{alg1} terminates after a finite number $\breve{n}$ of iterations and $\mathbf{x}^{(\breve{n})}$ is a stationary point of \eqref{sec4_3};
		\item[\emph{2)}] Algorithm~\ref{alg1} generates an infinite sequence $\left\{\mathbf{x}^{(n)}\right\}$ and
%If either one of the following conditions is satisfied:
%		\begin{enumerate}
%			\item[\emph{i)}] Either $f_0(\cdot)$ or $g_0(\cdot)$ is strongly convex
%			\item[\emph{ii)}] There exists an index $i_0$ such that either $\pmb{\mathcal{F}}_{i_0}(\cdot)$ or $\pmb{\mathcal{G}}_{i_0}(\cdot)$ is strongly convex\footnote{We extend the definition of the strong convexity of a scalar function to PSD-convex mappings. Specifically, a PSD-convex mapping $\pmb{\mathcal{F}}(\cdot)$ is said strongly convex with parameter $\rho>0$ if the following inequality holds for all $\mathbf{x},\mathbf{y}\in\Omega$: $\left(\pmb{\mathcal{DF}}(\mathbf{x})-\pmb{\mathcal{DF}}(\mathbf{y})\right)\left(\mathbf{x}-\mathbf{y}\right)\succeq\frac{\rho}{2n}\lVert\mathbf{x}-\mathbf{y}\rVert_2^2\mathbf{I}_n$, or equivalently, $\pmb{\mathcal{F}}(\mathbf{y})\succeq \pmb{\mathcal{F}}(\mathbf{x})+\pmb{\mathcal{DF}}(\mathbf{x})(\mathbf{y}-\mathbf{x})+\frac{\rho}{2n}\lVert\mathbf{x}-\mathbf{y}\rVert_2^2\mathbf{I}_n$.}. and $\bar{\pmb{\Phi}}_{i_0}\triangleq\lim_{n\rightarrow}\pmb{\Phi}_{i_0}^{(n)}\neq\mathbf{0}$, where $\pmb{\Phi}_{i_0}^{(n)}$ denotes the dual variable associated with the $i$\textsuperscript{th} constraint in \eqref{sec4_52} at the $n$\textsuperscript{th} iteration,
%		\end{enumerate}
		any limit point of this sequence is a stationary point of \eqref{sec4_3}.
	\end{itemize}
\end{theorem}
\begin{IEEEproof}
	Please see Appendix~\ref{appendix2}.
\end{IEEEproof}

Based on the above results, it is not difficult to further prove that the sequence of the objective function  $\left\{\varphi(\xn)\right\}$ of \eqref{sec4_3} generated by Algorithm~\ref{alg1} is also convergent.

\section{Sub-Optimal Solution to the Robust Secrecy Problem Based on the Penalized DC Algorithm}\label{sec5}
In this section, we develop a solution to our robust secure relaying problem \eqref{sec2_11} by applying the proposed penalized DC algorithmic framework.

\subsection{Reformulating \eqref{sec2_11} as a PSD-DC Program}
The transformation of \eqref{sec2_11} into a standard PSD-DC program as defined in \eqref{sec4_3}, involves several changes of variables and matrix manipulations. Specifically, by introducing a new variable  $\mathbf{U}\triangleq\sigma_{\Ss}\mathbf{W}$ and using \eqref{sec3_4}, \eqref{sec2_11} can be reformulated as follows:
\begin{subequations}\label{sec5_1}
	\begin{align}
		\label{sec5_11}
		\min_{\substack{\mathbf{W},\pmb{\Psi},\sigma_{\Ss}\\\mathbf{U},t,\pmb{\rho}}}&\quad -|\mathbf{h}_2^H\mathbf{U}\mathbf{h}_1|^2/t\\
		\label{sec5_12}
		&\quad\sigma_{\R}^2\lVert\mathbf{h}_2^H\mathbf{W}\rVert^2+\tr(\mathbf{h}_2\mathbf{h}_2^H\pmb{\Psi})+\sigma_{\D}^2\leq t\\
        \label{sec5_13}
		&\quad\lVert\mathbf{U}\mathbf{h}_1\rVert^2+\sigma_{\R}^2\lVert\mathbf{W}\rVert_F^2+\tr(\pmb{\Psi})\leq P_{\R}\\
		\label{sec5_14}
		&\quad 0\leq\sigma_{\Ss}\leq\bar{\sigma_{\Ss}},~\pmb{\Psi}\succeq\mathbf{0}\\
		 &\quad\underbrace{\mathbf{P}_k^H\mathbf{U}\mathbf{h}_1\mathbf{h}_1^H\mathbf{U}^H\mathbf{P}_k-\pmb{\Lambda}_k(\rho_k)-\gamma\mathbf{P}_k^H\pmb{\Psi}\mathbf{P}_k}_{\F_k(\mathbf{U},\pmb{\Psi},\pmb{\rho})}\nonumber\\
		&\qquad\qquad-\underbrace{\gamma\sigma_{\R}^2\mathbf{P}_k^H\mathbf{W}\mathbf{W}^H\mathbf{P}_k}_{\G_k(\mathbf{W})}\preceq\mathbf{0},~k\in\K\\
		\label{sec5_16}
		&\quad\mathbf{U}=\sigma_{\Ss}\mathbf{W},
	\end{align}
\end{subequations}
where $t$ is an auxiliary variable and we recall that $\bar{\sigma}_{\Ss}$ is the upper-bound of $\sigma_{\Ss}$, as defined in \eqref{sec3_9}. It is not difficult to verify that both functions $\F_k(\cdot)$ and $\G_k(\cdot)$ defined above are PSD-convex mappings. The above problem is still non-convex due to the bilinear equality constraint \eqref{sec5_16}. To this end, we rely on \cite[Lemma 1]{Yang_ICC} for equivalently converting \eqref{sec5_16} to a combination of an LMI and a DC inequality:
%\begin{lemma}[Lemma~1 of \cite{Concave}]\label{lemma1}
%	Given $\mathbf{X}$, $\mathbf{A}$, $\mathbf{B}$ and $\mathbf{C}$ of appropriate dimensions, which satisfy
%	\begin{equation}\label{e57}
%	\mathbf{X}=\mathbf{A}\mathbf{B}\mathbf{C}
%	\end{equation}
%	then the matrix equality \eqref{e57} is equivalent to the following inequality constraints:
%	\begin{align}
%	\left[\begin{array}{ccc}
%	\pmb{\mathcal{L}}_{11} & \mathbf{X} & \mathbf{A}\mathbf{B}\\
%	\mathbf{X}^H & \pmb{\mathcal{L}}_{22} & \mathbf{C}^H\\
%	\mathbf{B}^H\mathbf{A}^H & \mathbf{C} & \mathbf{I}
%	\end{array}\right]&\succeq\mathbf{0}\\
%	\tr\left(\pmb{\mathcal{L}}_{11}-\mathbf{A}\mathbf{B}\mathbf{B}^H\mathbf{A}^H\right)&\leq0
%	\end{align}
%	where $\pmb{\mathcal{L}}_{11}$ and $\pmb{\mathcal{L}}_{22}$ are auxiliary matrix variables with appropriate dimensions.
%\end{lemma}
\begin{align}
	\label{sec5_2}
	&\left[
	\begin{array}{ccc}
		\pmb{\mathcal{L}}_{11} & \mathbf{U} & \sigma_{\Ss}\mathbf{I}_{N_{\R}}\\
		\mathbf{U}^H & \pmb{\mathcal{L}}_{22} & \mathbf{W}^H\\
		\sigma_{\Ss}\mathbf{I}_{N_{\R}} & \mathbf{W} & \mathbf{I}_{N_{\R}}
	\end{array}
	\right]\succeq\mathbf{0}\\
	\label{sec5_3}
	&\underbrace{\tr(\pmb{\mathcal{L}}_{11})}_{f_1(\pmb{\mathcal{L}}_{11})}-\underbrace{\tr(\sigma_{\Ss}^2\mathbf{I}_{N_{\R}})}_{g_1(\sigma_{
		\Ss})}\leq 0,
\end{align}
where $\pmb{\mathcal{L}}_{11}$ and $\pmb{\mathcal{L}}_{22}$ are auxiliary matrix variables. For notational simplicity, let us define  $\mathbf{x}=[\sigma_{\Ss},\vect(\mathbf{W})^T,\vect(\mathbf{U})^T,\vect(\pmb{\Psi})^T,t,\pmb{\rho}^T,\vect(\pmb{\mathcal{L}}_{11})^T,\vect(\pmb{\mathcal{L}}_{22})^T]^T$ collectively denoting all the design and auxiliary variables. The following compact convex set is also defined for ease of presentation:
\begin{align}
	\Omega=\{\mathbf{x}:\eqref{sec5_12},\eqref{sec5_13},\eqref{sec5_14},\eqref{sec5_2}\}.
\end{align}

Furthermore, in order to guarantee the strong convexity of the functions in the DC decomposition of \eqref{sec5_11}, we equivalently rewrite \eqref{sec5_11} as
\begin{align}
	 \varphi(\x)=\underbrace{\lVert\mathbf{x}\rVert^2}_{f_0(\cdot)}-\big(\underbrace{|\mathbf{h}_2^H\mathbf{U}\mathbf{h}_1|^2/t+\lVert\mathbf{x}\rVert^2}_{g_0(\cdot)}\big).
\end{align}
Then, using the above definitions,
\eqref{sec5_1} can be equivalently reformulated as a standard PSD-DC program:
\begin{subequations}\label{sec5_4}
	\begin{align}
		\min_{\mathbf{x}\in\Omega}&\quad
		f_0(\x)-g_0(\x)	
		\\
		\label{sec5_42}
		\st&\quad f_1(\pmb{\mathcal{L}}_{11})-g_1(\sigma_{\Ss})\leq 0\\
		\label{sec5_43}
		&\quad\pmb{\mathcal{F}}_k(\mathbf{U},\pmb{\Psi},\pmb{\rho})-\pmb{\mathcal{G}}_k(\mathbf{W})\preceq\mathbf{0},~k\in\K,
	\end{align}
\end{subequations}
which completes the transformation.

Before proceeding further to the algorithm design, a key observation is of interest, which essentially motivates the development of the penalized DC framework:
\begin{remark}[On the Initialization of \eqref{sec5_4}]
	As mentioned in Section~\ref{sec4}, solving \eqref{sec5_4} by the conventional DC algorithm would require finding a feasible initialization, which corresponds to the non-convex feasibility check problem:
	\begin{align*}
		\texttt{Find}~\mathbf{x}\in\Omega~~\st~ \eqref{sec5_42},\eqref{sec5_43}.
	\end{align*}
	However, solving the above problem requires the same amount of computational efforts as solving \eqref{sec5_4}. If we turn our attention back to the original formulation of \eqref{sec2_11}, the only trivial initialization can be $(\sigma_{\Ss},\mathbf{W},\pmb{\Psi})=(0,\mathbf{0},\mathbf{0})$, which corresponds to the ``silent mode'' with zero information leakage and hence satisfies the robust secrecy constraints \eqref{sec2_112}. Unfortunately, a careful inspection reveals that $(0,\mathbf{0},\mathbf{0})$ is already a stationary point of \eqref{sec5_4} and clearly the conventional DC algorithm will stall at the point $(0,\mathbf{0},\mathbf{0})$, which is not a meaningful solution from a practical perspective.
\end{remark}

\subsection{Algorithm Design}
To involve $\algone$ for solving \eqref{sec5_4}, now only an initialization $\x^{(0)}\in\Omega$ is required. Since $\Omega$ is convex and compact, we are able to exploit its structure, i.e., \eqref{sec5_12}--\eqref{sec5_14}, and conveniently select a reasonably ``good'' starting point as follows:
$\sigma_{\Ss}^{(0)}=\bar{\sigma_{\Ss}}$, $\pmb{\Psi}^{(0)}=\mathbf{0}$,
$\mathbf{W}^{(0)}=\left(\frac{P_{\R}}{\bar{\sigma_{\Ss}}^2\lVert\mathbf{h}_1\rVert^2+\sigma_{\R}^2N_{\R}}\right)^{\frac{1}{2}}\mathbf{I}_{N_{\R}}$ and $\mathbf{U}^{(0)}=\bar{\sigma_{\Ss}}\mathbf{W}^{(0)}$.
To construct the algorithm, we need to linearize the concave parts $-g_0(\cdot)$, $-g_1(\cdot)$ and $-\G_k(\cdot)$ by their first-order Taylor series expansion, which corresponds to Step 1 in $\algone$ and leads to:
\begin{align}
	&-\hat{g}_0(\x;\xn)\nonumber\\
	 &\qquad=-\frac{\left|\mathbf{h}_2^H\mathbf{U}^{(n)}\mathbf{h}_1\right|^2}{t^{(n)}}+\frac{\left|\mathbf{h}_2^H\mathbf{U}^{(n)}\mathbf{h}_1\right|^2}{(t^{(n)})^2}\left(t-t^{(n)}\right)\nonumber\\
	&\qquad\quad~ -\frac{1}{t^{(n)}}2\rea\left\{\mathbf{h}_2^H\mathbf{U}^{(n)}\mathbf{h}_1\mathbf{h}_1^H\left(\mathbf{U}-\mathbf{U}^{(n)}\right)^H\mathbf{h}_2\right\}\nonumber\\
	&\qquad\quad~-\xn\big(\x-\xn\big)^H\\
	&-\hat{g}_1(\sigma_{\Ss};\sigma_{\Ss}^{(n)})=-N_{\R}(\sigma_{\Ss}^{(n)})^2-2N_{\R}\sigma_{\Ss}^{(n)}\left(\sigma_{\Ss}-\sigma_{\Ss}^{(n)}\right)\\
	&-\hat{\G}_k(\mathbf{W};\mathbf{W}^{(n)})=-\gamma\sigma_{\R}^2\mathbf{P}_k^H\mathbf{W}^{(n)}\left(\mathbf{W}^{(n)}\right)^H\mathbf{P}_k\nonumber\\
	&\qquad\qquad\qquad\quad-2\gamma N_{\R}\rea\left(\mathbf{P}_k^H\mathbf{W}^{(n)}\left(\mathbf{W}-\mathbf{W}^{(n)}\right)^H\mathbf{P}_k\right).
\end{align}
It is observed that $\Omega$ can be represented by a set of LMIs based on the technique introduced in \cite{SOCP} (details of this transformations are omitted for brevity), and $\F_k(\cdot)$ is Schur PSD-convex. Then, using the technique discussed under Remark~1-4), the penalized version of \eqref{sec5_4} can be recast as the following standard SDP:
%\footnote{In implmentation, we add two constraints to $\Omega$, namely $\sigma_{\Ss}\geq\epsilon, \mathbf{W}\succeq\epsilon\mathbf{I}_{N_{\R}}$, to avoid the case where the algorithm converges to the stationary point of $(\sigma_{\Ss}=0, \mathbf{W}=\mathbf{0})$.}
\begin{align}\label{sec5_5}
	\min_{\substack{\mathbf{x}\in\Omega,s\\
			\{\mathbf{T}_k\},\{\mathbf{S}_k\}}}&\quad f_0(\x)-\hat{g}_0(\x;\xn)+\tau^{(n)}\big(s+\sum_{k=1}^K\tr(\mathbf{S}_k)\big)\nonumber\\
	\st&\quad f_1(\pmb{\mathcal{L}}_{11})-\hat{g}_1(\sigma_{\Ss};\sigma_{\Ss}^{(n)})\leq s\nonumber\\
	&\quad\mathbf{T}_k-\pmb{\Lambda}(\rho_k)-\gamma\mathbf{P}_k^H\pmb{\Psi}\mathbf{P}_k-\hat{\G}_k(\mathbf{W};\mathbf{W}^{(n)})\preceq\mathbf{S}_k\nonumber\\
	&\quad\left[
		\begin{array}{cc}
			\sigma_{\R}^2\mathbf{I}_{N_{\R}} & \mathbf{h}_1^H\mathbf{U}^H\mathbf{P}_k\\ \mathbf{P}_k^H\mathbf{U}\mathbf{h}_1 & \mathbf{T}_k
		\end{array}
	\right]\succeq\mathbf{0}\nonumber\\
	&\quad s\geq 0,~\mathbf{S}_k\succeq\mathbf{0},~k\in\K,
\end{align}
where $s$ and $\{\mathbf{S}_k\}$ are the feasibility indicators, and $\{\mathbf{T}_k\}$ represents a set of auxiliary variables. Subsequently, replacing \eqref{sec4_5} with the above specified sub-problem in $\algone$, we arrive at the penalized DC algorithm for \eqref{sec5_4} with proven convergence towards a stationary point\footnote{It is not difficult verify that A.1)--A.3) hold for \eqref{sec5_4}, and therefore, the convergence result in Theorem~\ref{theorem1} is valid for \eqref{sec4_5}.}.

\section{Numerical Simulations}\label{sec6}
The efficacy of the proposed solutions to the robust secure relaying problem is verified by numerical simulations. In our experiments, the coefficients of the legitimate channels $\mathbf{h}_1$ and $\mathbf{h}_2$, as well as the estimated $\eve$s' channels $\{\hat{g}_{1,k}\}$ and $\{\hat{\mathbf{g}}_{2,k}\}$ are generated by identically and independently distributed (i.i.d.) complex circular Gaussian distributions with zero-mean and unit-variance. Equal radii are assumed for all $\Delta g_{1,k}$ and for all $\Delta \mathbf{g}_{2,k}$, i.e., $\varepsilon_{1,k}=\varepsilon_1$ and $\varepsilon_{2,k}=\varepsilon_2$ for all $k$. The power budget of $\Ss$ is normalized to $P_{\Ss}=1$ and that of $\R$ is set to $P_{\R}=2$. It is also assumed that an antenna array of size $N_{\R}=3$ is employed by $\R$. The noise variances are set to $\sigma_{\R}^2=0.05$, $\sigma_{\D}^2=0.05$ and $\sigma_{\E,k}^2=0.01~\forall k$. The above parameters are fixed, unless otherwise stated.

\begin{figure}[t]
	\centering
	\includegraphics[width=9cm]{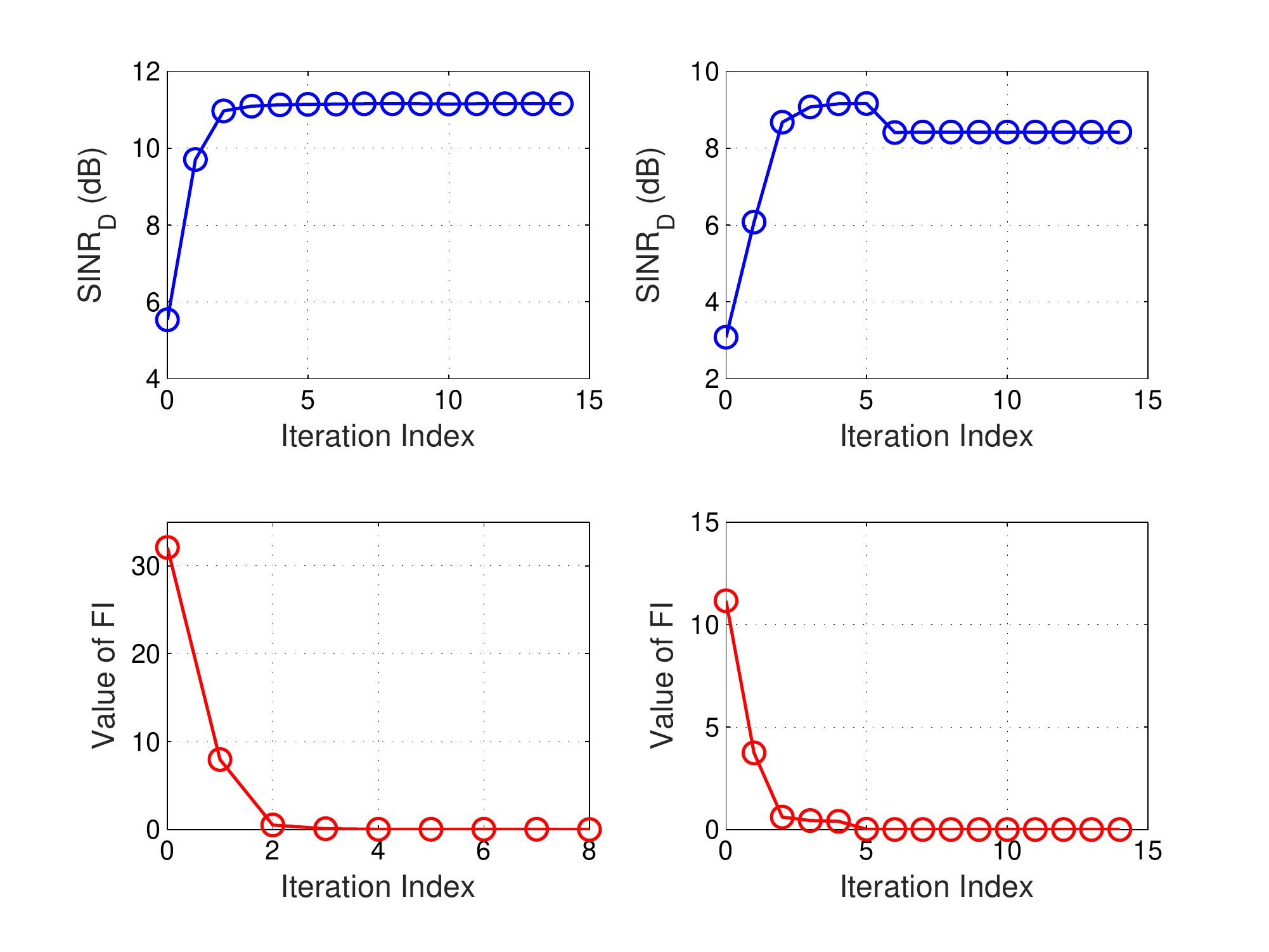}\\
	\caption{Convergence behavior of $\algone$. Left: Case 1. Right: Case 2.}\label{fig2}
	\includegraphics[width=9cm]{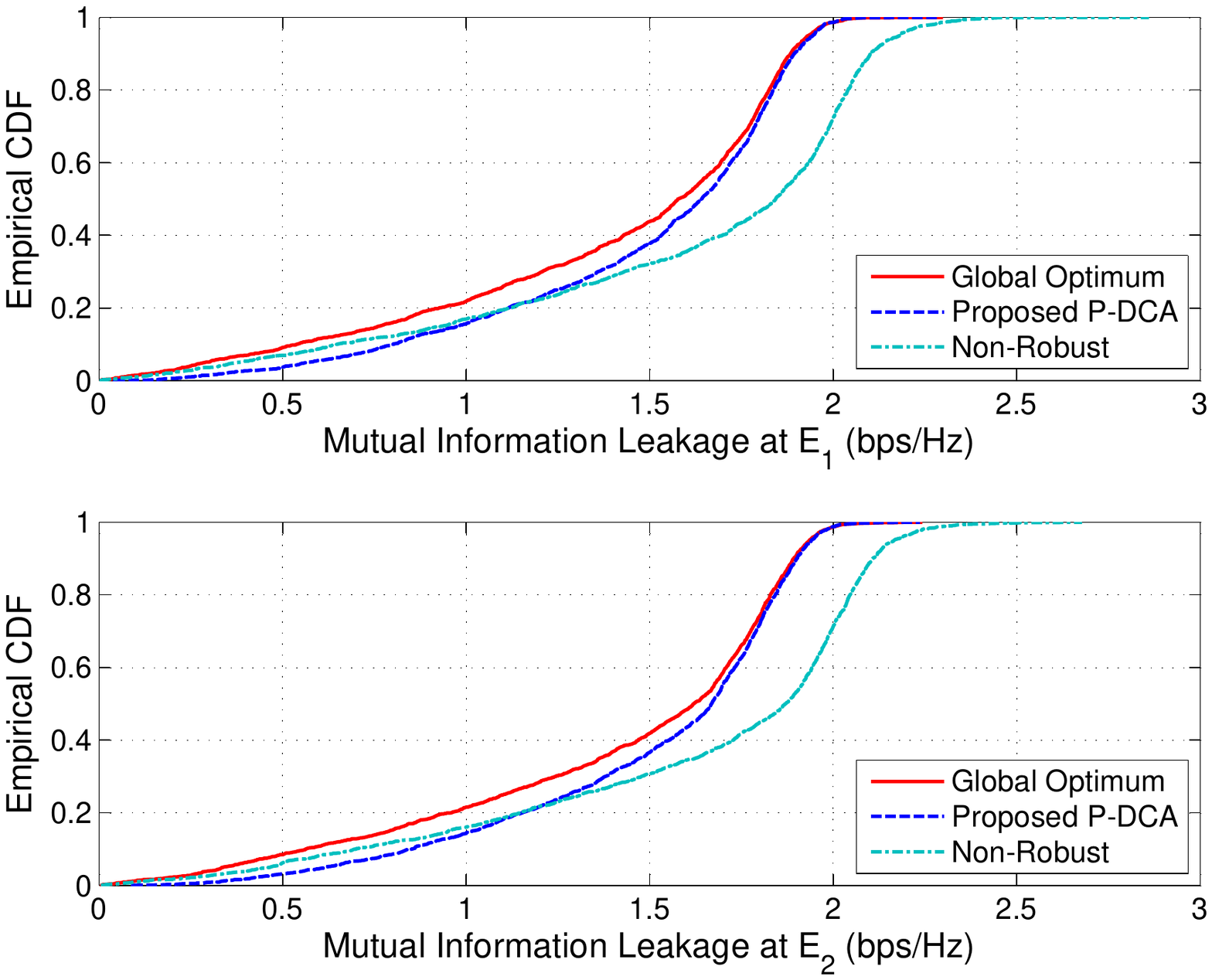}\\
	\caption{Empirical CDFs of the mutual information leakage at $\eve$s. The legitimate $\Ss$ is transmitting at $R_d=2$ bps/Hz.}\label{fig3}
\end{figure}
\subsubsection{Convergence}
We first study the convergence behavior of $\algone$. We simulate $200$ channel realizations and among these, two different types of behaviors are observed; a representative case for each type is then plotted in the left and right parts of Fig.~\ref{fig2}. In each case, the top sub-figure shows the convergence of the SINR achieved at $\D$, whilst the bottom sub-figure plots the evolution of the feasibility indicator. The first case (left) exhibits a behavior similar to the conventional DC algorithm. The second case (right) shows a more interesting behavior, where the algorithm begins with an infeasible point and attempts to find a region (still infeasible) associated with a larger objective value. As the penalty term in \eqref{sec4_51} begins to play a gradually more important role, the emphasis shifts towards finding a feasible point near the above located region. Hence, the value of SINR drops, since the feasibility has to be enforced now. Finally, the SINR remains approximately constant, because a stationary point has been reached. The convergence behavior is consistent with the discussions and proof provided in Section~\ref{sec4}.

\subsubsection{Secrecy}
To evaluate the relay-aided transmission secrecy achieved by the proposed solutions, i.e., to quantify how consistently the robust secrecy constraints \eqref{sec2_112} can be satisfied, we follow a probabilistic approach similar to \cite[Section VI--B]{Yang_TVT}. In this example, the coefficients of $\Delta\hat{g}_{1k}$ and $\Delta\hat{\mathbf{g}}_{2k}$ are generated as i.i.d. zero-mean complex circular Gaussian random variables with a variance of $\sigma_{h}^2=0.05$. The radii of the uncertainty regions in \eqref{sec2_10} and \eqref{sec2_12} are then determined by $\varepsilon_1=\sigma_{h}^2\mathtt{gammaincinv}(\text{Pr},0.5)$ and $\varepsilon_2=\sigma_{h}^2\mathtt{gammaincinv}(\text{Pr},0.5N_{\R}^2)$, where $\mathtt{gammaincinv}(\cdot)$ represents the inverse incomplete Gamma function, as defined in \texttt{MATLAB} and $\text{Pr}$ is a predefined bounding probability, say, $\text{Pr}=95\%$, c.f.\cite[(61)]{Yang_TVT}. The empirical cumulative distribution functions (CDFs) of the mutual information leakage at both $\eve$s are shown in Fig~\ref{fig3}. Both the proposed solutions ensure that the mutual information leakage stays below the data rate of legitimate UEs for more than $98\%$ of the realizations, whilst the non-robust design leads to a frequent violation of the secrecy constraints, namely for more than $25\%$ of the realizations.
\subsubsection{Reliability}
\begin{figure}[t]
	\centering
	\includegraphics[width=9cm]{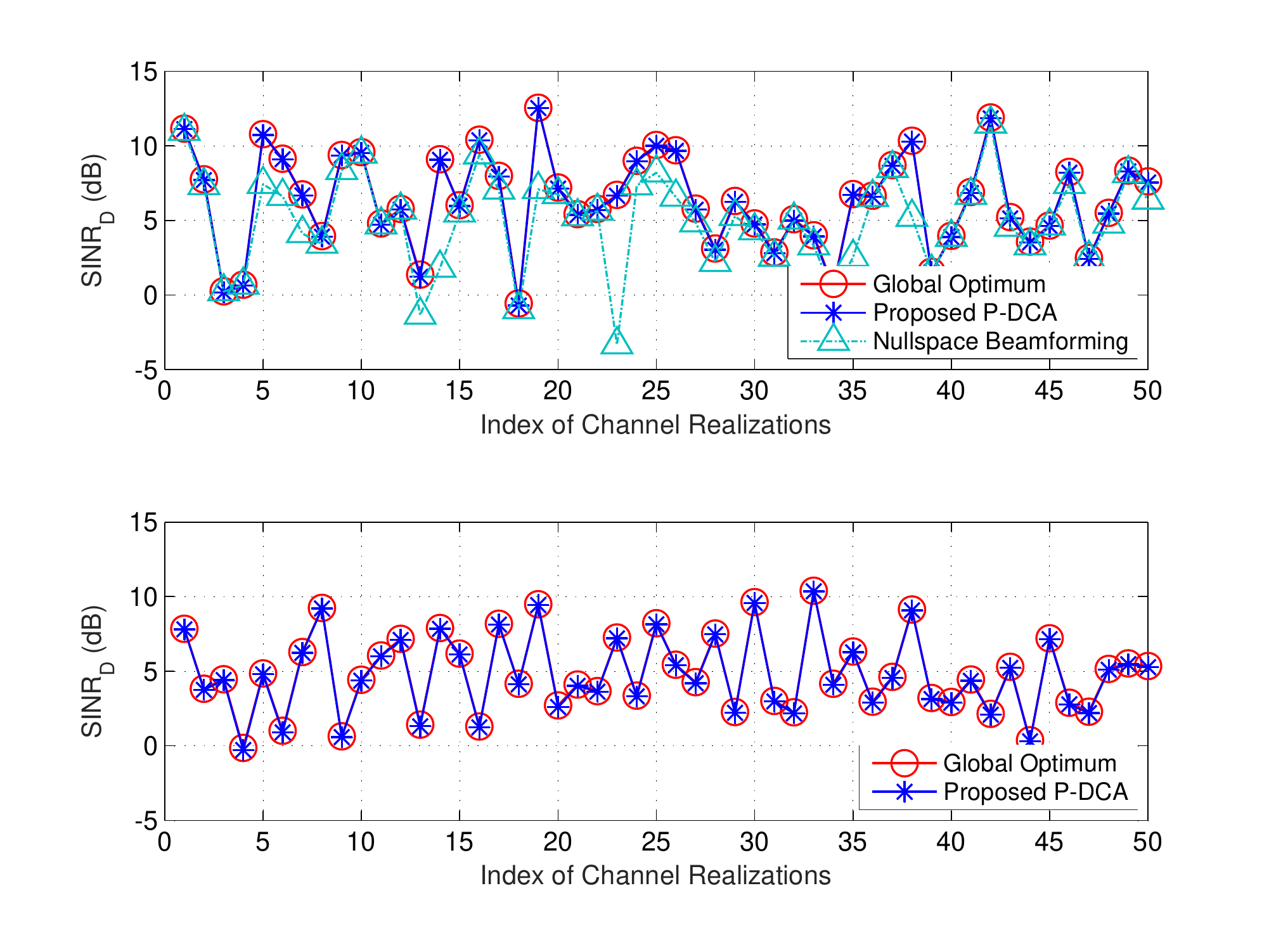}\\
	\caption{Achieved SINR at $\D$. Top sub-figure: $K=2$ $\eve$s. Bottom sub-figure: $K=4$ $\eve$s.}\label{fig4}
	\includegraphics[width=9cm]{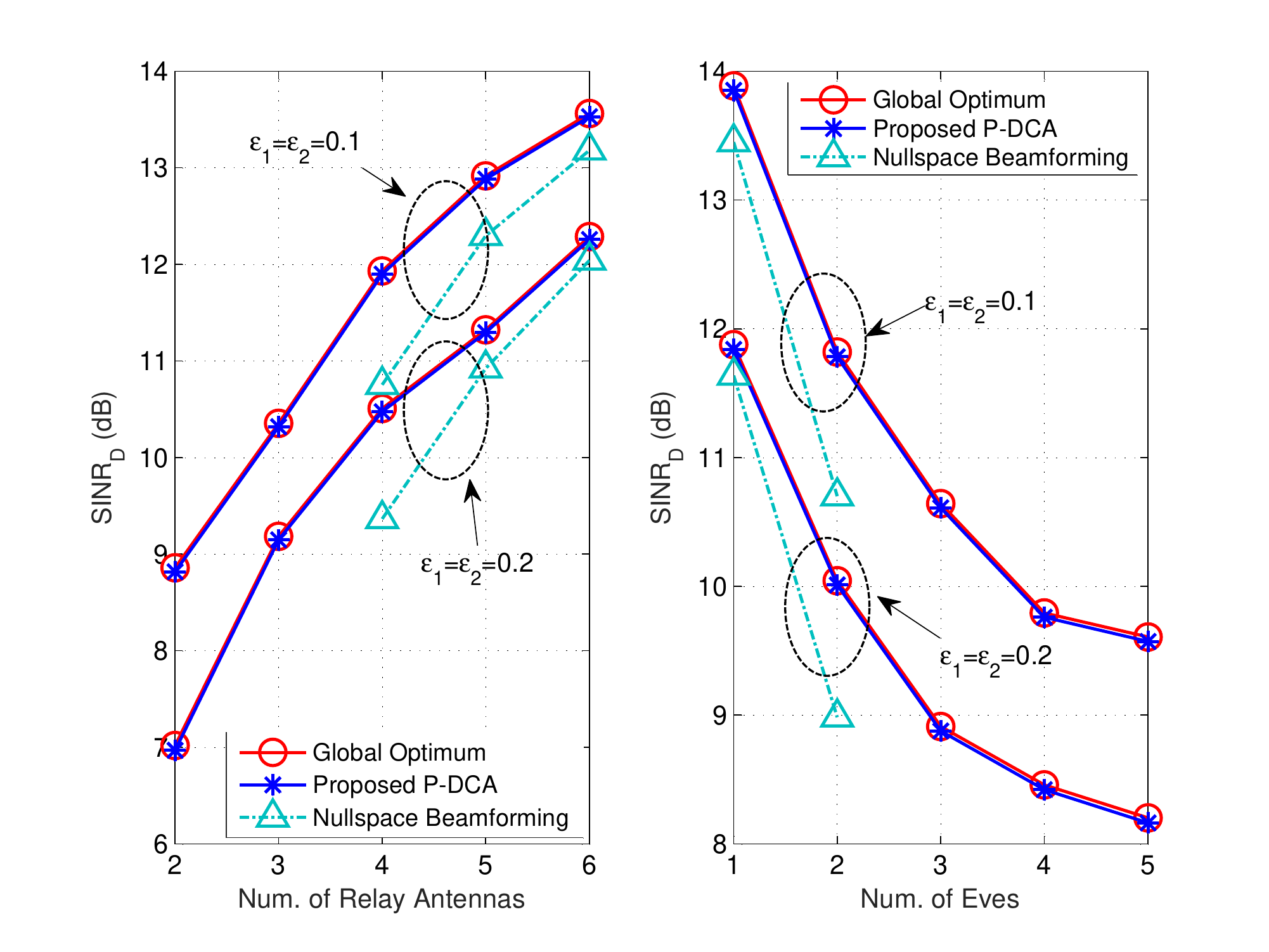}\\
	\caption{Achieved SINR at $\D$. Left sub-figure: $\SINR_{\D}$ versus $N_{\R}$. $K=3$ $\eve$s are considered. Right sub-figure: $\SINR_{\D}$ versus $K$. }\label{fig5}
\end{figure}
Having verified the secrecy of the proposed solutions, we now compare the attainable transmission reliability in terms of the SINR achieved at $\D$. In Fig.~\ref{fig4}, $\SINR_{\D}$ is plotted for a set of $50$ independent experiments and two different numbers of $\eve$s, i.e., $K=2$ and $K=4$, are considered. The curve labeled ``Nullspace Beamforming'' refers to a method, where $\R$ attempts to nullifies $\eve$s' reception by exploiting the null space of $[\hat{\mathbf{g}}_{2,1},\cdots,\hat{\mathbf{g}}_{2,K}]$ in the design of the AF relaying matrix \cite{Relay_Beamforming,Robust_AF_AN}. Therefore, the method is only applicable, when $N_{\R}>K$. In both cases, we observe that the performance of the proposed penalized DC algorithm is very close to the globally optimal solution. In the case of $K=2$, both the proposed solutions significantly outperform the conventional nullspace beamforming method.

\begin{figure}[t]
	\centering
	\includegraphics[width=9cm]{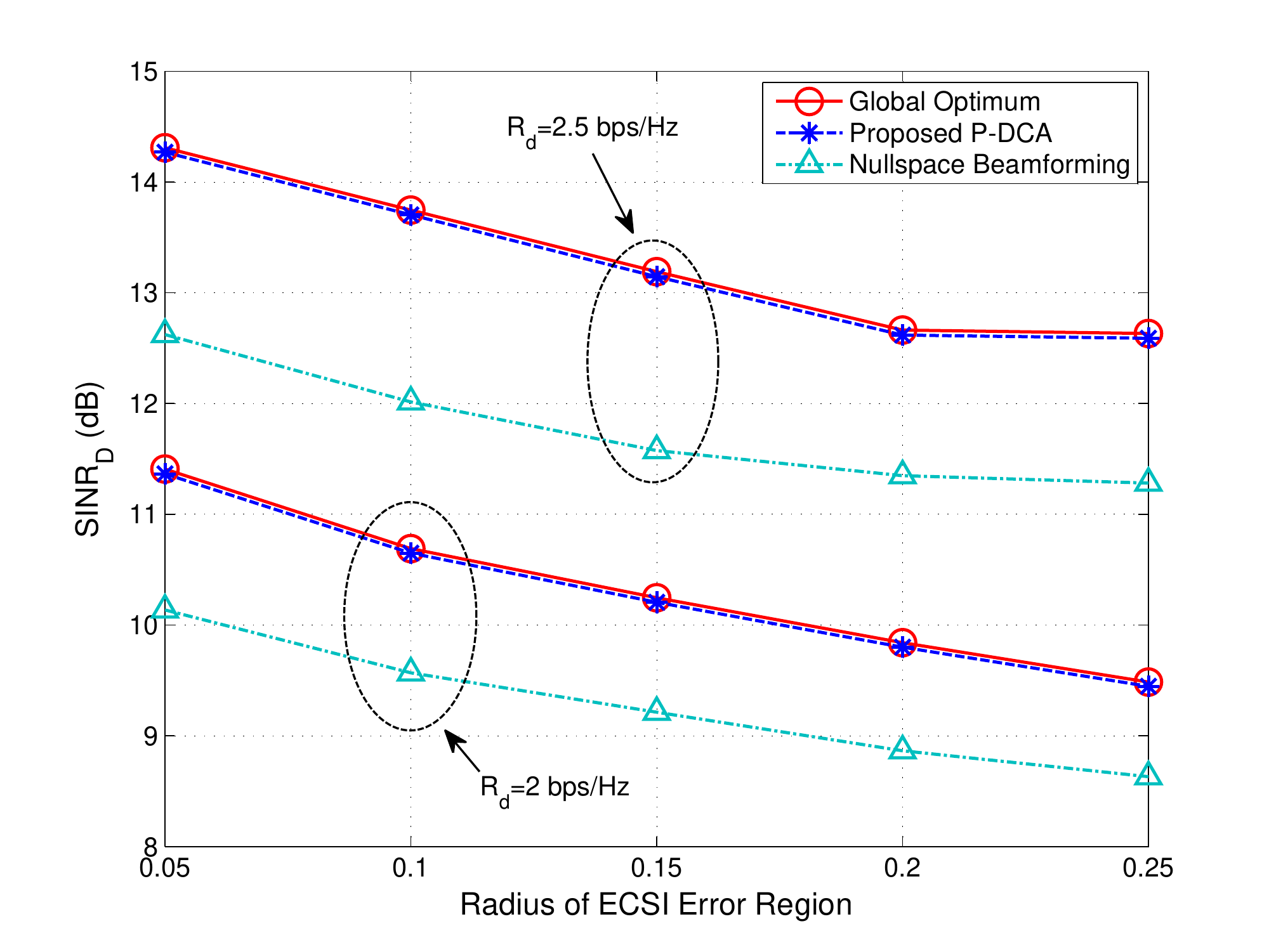}\\
	\caption{Achieved SINR at $\D$ as a function of size of uncertainty region. Two data rates of legitimate UEs are considered, namely, $R_d=2$bps/Hz and $R_d=2.5$bps/Hz.}\label{fig6}
\end{figure}

We then study how different system configurations impact the SINR achieved by the different approaches. In the left sub-figure of Fig.~\ref{fig5}, the  SINR achieved by the proposed solutions and by the conventional nullspace beamforming is plotted as a function of the number of antenna elements employed at $\R$ in the case of $K=3$. Two different sizes of uncertainty regions are considered, i.e., $\varepsilon_1=\varepsilon_2=0.1$ and $\varepsilon_1=\varepsilon_2=0.2$. The SINR achieved monotonically increases with $N_{\R}$ in both cases  due to the higher diversity gleaned from the antenna array. Again, both the proposed solutions consistently exhibit a better performance than the conventional nullspace beamforming. Observe that when the channel uncertainty is increased ($\varepsilon_1=\varepsilon_2=0.2$), the legitimate UEs are confined to a relatively low transmission power for satisfying the robust secrecy constraints, leading to a lower received SINR at $\D$. In the right sub-figure of Fig.~\ref{fig5}, the impact of different number of $\eve$s on the SINR achieved is assessed in the case of $N_{\R}=3$. The SINR monotonically decreases, when there are more $\eve$s and therefore, the legitimate UEs have to reduce their transmission power for preventing the information leakage. For completeness, we also investigate how robustly the proposed solutions can behave against the ECSI errors by varying the size of the uncertainty regions over a wider range. The results, which are shown in Fig.~\ref{fig6} again demonstrates the superiority of our proposed solutions.
\subsubsection{Computational Complexity}
\begin{table}[t]
	\caption{Average solver time (in seconds) for different algorithms}\label{t1}
	\centering
	\begin{tabular}{|c|c|c|c|c|c|c|c|}
		\cline{4-8}
		\multicolumn{2}{c}{} & {} & \multicolumn{5}{c|}{Num. of Relay Ant. $N_{\R}$} \\
		\cline{3-8}
		\multicolumn{2}{c|}{} & Alg. & $2$ & $3$ & $4$ & $5$ & $6$\\
		\hline
		\multirow{8}{*}{\begin{sideways}Num. of $\eve$s $K$\end{sideways}} &
		\multirow{2}{*}{$2$} & Global & $1.31$ & $4.44$ & $22.44$ & $111.26$ & $553.11$\\
		{} & {} & P-DCA & $0.69$ & $1.42$ & $2.58$ & $4.73$ & $8.38$\\
		%{} & {} & (In)P-DCA & $0.22$ & $0.32$ & $0.52$ & $0.77$ & $1.06$ \\
		\cline{2-8}
		{} & \multirow{2}{*}{$3$} & Global & $1.59$ & $5.95$ & $28.61$ & $141.54$ & $680.90$\\
		{} & {} & P-DCA & $0.88$ & $1.80$ & $3.45$ & $6.52$ & $11.31$\\
		%{} & {} & (In)P-DCA & $0.28$ & $0.51$ & $0.66$ & $1.15$ & $1.78$ \\
		\cline{2-8}
		{} & \multirow{2}{*}{$4$} & Global & $1.83$ & $7.14$ & $33.29$ & $165.29$ & $798.69$\\
		{} & {} & P-DCA & $1.05$ & $2.24$ & $4.60$ & $8.01$ & $14.09$\\
		%{} & {} & (In)P-DCA & $0.34$ & $0.55$ & $0.97$ & $1.31$ & $2.26$\\
		\cline{2-8}
		{} & \multirow{2}{*}{$5$} & Global & $2.19$ & $8.38$ & $41.23$ & $203.20$ & $924.32$\\
		{} & {} & P-DCA & $1.20$ & $2.66$ & $5.43$ & $9.95$ &  $17.12$\\
		%{} & {} & (In)P-DCA & $0.40$ & $0.66$ & $1.49$ & $1.99$ & $3.09$\\
		\hline
	\end{tabular}
\end{table}
Last but not least, we have to evidence the lower complexity of the proposed penalized DC algorithm as compared to the globally optimal solution. The solver time of the different algorithms averaged over $100$ independent realizations is shown in Table~\ref{t1} for different values of $N_{\R}$ and $K$. It is observed that the solver time of the globally optimal solution escalates rapidly upon increasing  $N_{\R}$ or $K$, which is consistent with the worst-case complexity analysis of Section~\ref{sec3}. By contrast, the solver time of the proposed penalized DC algorithm increases more slowly compared to the former. Specifically, when the number of relay antennas $N_{\R}$ becomes large, the solver time of the penalized DC algorithm is less than $5\%$ of that of the globally optimal solution, which demonstrates the gains of the proposed algorithm in terms of computational complexity.

%Furthermore, with the aid of inexact implementation of the DC algorithm\footnote{To adapt Algorithm~\ref{alg1}(In) to \eqref{sec5_5}, we follow the inspiration of \cite{SCA}. Specifically, at each iteration of Algorithm~\ref{alg1}(In), instead of exactly solving \eqref{sec5_5}, we apply the so-called block coordinate descent (BCD)-type update by dividing the variables into two blocks, namely, $\mathcal{X}_1=\mathbf{x}\backslash\{\sigma_{\Ss}\}$ and $\mathcal{X}_2=\sigma_{\Ss}$ and perform one round of BCD-type update of \eqref{sec5_5}, i.e., solve \eqref{sec5_5} with respect to each of $\mathcal{X}_1$ and $\mathcal{X}_2$ only once. It is explicitly shown in \cite{SCA} that the above implementation is a special case of the inexact implementation proposed in Section~\ref{sec4}-C and satisfies C1) and C2). Therefore, the convergence results in  Theorem~\ref{theorem2} apply.}, the solver time is further reduced by more than $60\%$, demonstrating the gains of incorporating the concept of inexact optimization into the proposed penalized DC framework.

\section{Conclusions}\label{sec7}
Robust design of secure MIMO relaying was studied in the presence of multiple $\eve$s. We jointly optimized the power of $\Ss$, the AF matrix and the AN covariance at $\R$ for maximizing the received SINR at $\D$, while imposing a set of mutual information leakage-based secrecy constraints. Given only imperfect ECSI, the resultant problem has been shown to be non-convex and challenging. First a globally optimal solution based on SDR was proposed. To alleviate the high complexity associated with this method, a computationally efficient sub-optimal solution relying on a newly proposed penalized DC algorithm was developed. This algorithm is capable of efficiently finding a stationary solution to a general non-convex SDP represented by a PSD-DC program, without the need of a non-trivial feasible initialization.
%An inexact version of this algorithm was also proposed to further reduce the computational efforts.
We benchmarked the performance of the proposed solutions using a few numerical examples. It shows that the proposed solutions yield a significantly better performance than the non-robust and null-space beamforming methods. Additionally, the penalized DC algorithm typically attains a performance close to that of the globally optimal solution.

\appendices
\section{Proof of Proposition~\ref{proposition1}}\label{appendix1}
Firstly, expanding all the quadratic terms in $\mathbf{W}$ in \eqref{sec3_51} and \eqref{sec3_52}, and invoking the identities $\tr\left(\mathbf{A}^H\mathbf{B}\mathbf{C}\mathbf{D}^H\right)=\vect\left(\mathbf{A}\right)^H\left(\mathbf{D}^T\otimes\mathbf{B}\right)\vect\left(\mathbf{C}\right)$ and $\tr\left(\mathbf{A}^H\mathbf{B}\mathbf{A}\right)=\vect\left(\mathbf{A}\right)^H\left(\mathbf{I}\otimes\mathbf{B}\right)\vect\left(\mathbf{A}\right)$, \eqref{sec3_51} and \eqref{sec3_52} can be recast as \eqref{sec3_61} and \eqref{sec3_62}, respectively, with $\mathbf{X}=\mathbf{w}\mathbf{w}^H$.

Next, we transform \eqref{sec3_53}. Recall that $\pmb{\Theta}(\mathbf{W},\pmb{\Psi})=\sigma_{\Ss}^2\mathbf{W}\mathbf{h}_1\mathbf{h}_1^H\mathbf{W}^H-\gamma\sigma_{\R}^2\mathbf{W}\mathbf{W}^H-\gamma\pmb{\Psi}$, where the first term on the right hand side is equivalent to
\begin{equation}\label{a1}
\sigma_{\Ss}^2\mathbf{W}\mathbf{h}_1\mathbf{h}_1^H\mathbf{W}^H=\sigma_{\Ss}^2(\mathbf{h}_1^T\otimes\mathbf{I}_{N_{\R}})\mathbf{w}\mathbf{w}^H(\mathbf{h}_1^T\otimes\mathbf{I}_{N_{\R}})^H,
\end{equation}
which follows by using $\vect(\mathbf{A}\mathbf{B}\mathbf{C})=(\mathbf{C}^T\otimes\mathbf{A})\vect(\mathbf{B})$. To transform the second term, we write $\mathbf{W}=\left[\pmb{w}_1,\cdots,\pmb{w}_l,\cdots,\pmb{w}_{N_{\R}}\right]$, where $\pmb{w}_l$ denotes the $l$\textsuperscript{th} column of $\mathbf{W}$.
Then $\mathbf{W}\mathbf{W}^H$ can be equivalently expressed as
\begin{equation}\label{a4}
\mathbf{W}\mathbf{W}^H=\sum_{l=1}^{N_{\R}}\pmb{w}_l\pmb{w}_l^H.
\end{equation}
By expressing the linear relation between $\pmb{w}_l$ and $\mathbf{w}$ in a matrix form as $\pmb{w}_l=\mathbf{E}_l\mathbf{w}$, \eqref{a4} can further be written as
\begin{equation}\label{a2}
\mathbf{W}\mathbf{W}^H=\sum_{l=1}^{N_{\R}}\mathbf{E}_l\mathbf{w}\mathbf{w}^H\mathbf{E}_l^H.
\end{equation}
Using \eqref{a1} and \eqref{a2}, $\pmb{\Theta}(\mathbf{W},\pmb{\Psi})$ becomes equivalent to
\begin{equation}\label{a3}
\pmb{\Theta}(\mathbf{W},\pmb{\Psi})=\sigma_{\Ss}^2\mathbf{H}_1\mathbf{w}\mathbf{w}^H\mathbf{H}_1^H-\gamma\sigma_{\R}^2\sum_{l=1}^{N_{\R}}\mathbf{E}_l\mathbf{w}\mathbf{w}^H\mathbf{E}_l^H-\gamma\pmb{\Psi}.
\end{equation}
Finally, invoking $\mathbf{X}=\mathbf{w}\mathbf{w}^H$ and $\rank(\mathbf{X})=1$, \eqref{sec3_5} is readily reformulated as \eqref{sec3_6} with the aid of \eqref{a3}.

\section{Proof of Theorem~\ref{theorem1}}\label{appendix2}
We first have to examine the KKT conditions of the sub-problem \eqref{sec4_5}. Since the latter is convex and strictly feasible, i.e., Slater's constraint qualification is satisfied, the KKT sufficient conditions hold for the optimal solution $(\mathbf{x}^{(n+1)},\mathbf{S}^{(n+1)})$: there exist some dual variables  $\pmb{\Phi}_i^{(n+1)}\succeq\mathbf{0},\mathbf{Z}_i^{(n+1)}\succeq\mathbf{0}$ for $i\in\mathcal{I}$ so that \eqref{sec4_6}, shown at the top of the next page, holds.
\begin{figure*}
	\begin{subequations}\label{sec4_6}
		\begin{align}
			\label{sec4_61}
			\mathbf{0}\in\nabla f_0(\mathbf{x}^{(n+1)})-\nabla g_0(\mathbf{x}^{(n)})
			 +\sum_{i=1}^I\left(\pmb{\Phi}_i^{(n+1)}*\left(\pmb{\mathcal{DF}}_i(\mathbf{x}^{(n+1)})-\pmb{\mathcal{DG}}_i(\mathbf{x}^{(n)})\right)\right)+\mathcal{N}(\Omega,\mathbf{x}^{(n+1)})\\
			\label{sec4_62}
			\tau^{(n)}\mathbf{I}-\pmb{\Phi}_i^{(n+1)}-\mathbf{Z}_i^{(n+1)}=\mathbf{0},~i\in\mathcal{I}\\
			 \pmb{\mathcal{F}}_i(\mathbf{x}^{(n+1)})-\pmb{\mathcal{G}}_i(\mathbf{x}^{(n)})-\pmb{\mathcal{DG}}(\mathbf{x}^{(n)})(\mathbf{x}^{(n+1)}-\mathbf{x}^{(n)})\preceq\mathbf{S}_i^{(n+1)},~i\in\mathcal{I}\\
			\label{sec4_64}
			 \tr\left(\pmb{\Phi}_i^{(n+1)}\left(\pmb{\mathcal{F}}_i(\mathbf{x}^{(n+1)})-\pmb{\mathcal{G}}_i(\mathbf{x}^{(n)})-\pmb{\mathcal{DG}}(\mathbf{x}^{(n)})(\mathbf{x}^{(n+1)}-\mathbf{x}^{(n)})-\mathbf{S}^{(n+1)}_i\right)\right)=0,~i\in\mathcal{I}\\
			\label{sec4_65}
			 \mathbf{x}^{(n+1)}\in\Omega,~\mathbf{S}_i^{(n+1)}\succeq\mathbf{0},~\pmb{\Phi}_i^{(n+1)}\succeq\mathbf{0},~\mathbf{Z}_i^{(n+1)}\succeq\mathbf{0},~\tr(\mathbf{S}_i^{(n+1)}\mathbf{Z}_i^{(n+1)})=0,~i\in\mathcal{I}&
		\end{align}
	\end{subequations}
	\hrule
\end{figure*}
In \eqref{sec4_61}, $\mathcal{N}(\Omega,\mathbf{x})$ denotes the normal cone of $\Omega$ at $\mathbf{x}$, and ``$*$'' denotes a bilinear operator defined as $\mathbf{Z}*\mathbf{A}=[\tr(\mathbf{A}_1\mathbf{Z}),\cdots,\tr(\mathbf{A}_n\mathbf{Z})]^T$ for any $\mathbf{Z}\in\mathbb{H}^p$, where $\mathbf{A}=\sum_{i=1}^n x_i\mathbf{A}_i$ with $\mathbf{A}_i\in\mathbb{H}^p$ for $i=1,\cdots,n$.
	
We prove the first scenario. If Algorithm~\ref{alg1} terminates after a finite number $\breve{n}$ of iterations, it follows from the termination criterion (see Remark 1) that $\mathbf{x}^{(\breve{n}+1)}=\mathbf{x}^{(\breve{n})}$ and $\mathbf{S}_i^{(\breve{n}+1)}=\mathbf{0}$ for all $i$, i.e., $\breve{\mathbf{x}}$ is a feasible point of \eqref{sec4_3}. Letting $n=\breve{n}$ and substituting the above relations into \eqref{sec4_6}, we obtain \eqref{sec4_10}, as shown below \eqref{sec4_6}. A careful examination reveals the equivalence between \eqref{sec4_10} and the KKT conditions of \eqref{sec4_3}. Therefore, it is proved that $\mathbf{x}^{(\breve{n})}$ is a KKT stationary point of \eqref{sec4_3}.
\begin{figure*}	
	\begin{subequations}\label{sec4_10}
		\begin{align}
			\mathbf{0}\in\nabla f_0(\mathbf{x}^{(\breve{n})})-\nabla g_0(\mathbf{x}^{(\breve{n})})+\sum_{i=1}^I\left(\pmb{\Phi}_i^{(\breve{n})}*\left(\pmb{\mathcal{DF}}_i(\mathbf{x}^{(\breve{n})})-\pmb{\mathcal{DG}}_i(\mathbf{x}^{(\breve{n})})\right)\right)+\mathcal{N}(\Omega,\mathbf{x}^{(\breve{n})})&\\
			\pmb{\mathcal{F}}_i(\mathbf{x}^{(\breve{n})})-\pmb{\mathcal{G}}_i(\mathbf{x}^{(\breve{n})})\preceq\mathbf{0},~i\in\mathcal{I}&\\
			 \tr\left(\pmb{\Phi}_i^{(\breve{n}+1)}\left(\pmb{\mathcal{F}}_i(\mathbf{x}^{(\breve{n})})-\pmb{\mathcal{G}}_i(\mathbf{x}^{(\breve{n})})\right)\right)=0,~i\in\mathcal{I}&\\
			\mathbf{x}^{(\breve{n})}\in\Omega,~\pmb{\Phi}_i^{(\breve{n}+1)}\succeq\mathbf{0}.&
		\end{align}
	\end{subequations}
	\hrule
\end{figure*}
	
We now proceed to prove the second scenario. The key ingredients of the proof are to show that any limit point of $\{\xn\}$ is feasible to \eqref{sec4_3} and the sequence of the dual variable $\{\pmb{\Phi}_i^{(n)}\}$ is bounded, so that there exists at least one limit point of $\{\pmb{\Phi}_i^{(n)}\}$. Then we show that any primal-dual pair of the limit point satisfies the KKT conditions of \eqref{sec4_3}.
		
To prove that any limit point of $\{\xn\}$ is a feasible point of \eqref{sec4_3}, we rely on the following claims:
	\begin{claim}\label{claim1}
	There exists a finite iteration index $\tilde{n}$, so that
	\begin{equation}
		\tau^{(n)}=\tau^{(\tilde{n})},\quad \forall n\geq\tilde{n}.
	\end{equation}
\end{claim}
\begin{IEEEproof}
	Please see Appendix \ref{appendix3}.
\end{IEEEproof}
\begin{claim}\label{claim2}
	The sequence $\{\xn\}$ satisfies
	\begin{equation}\label{a2_2}
		\lim_{n\rightarrow\infty}\lVert\mathbf{x}^{(n+1)}-\mathbf{x}^{(n)}\rVert=0.
	\end{equation}
\end{claim}
\begin{IEEEproof}
	The proof is omitted here owing lack of space, but it can be found in the supplementary file uploaded with this manuscript.
\end{IEEEproof}
%\begin{claim}\label{claim2}
%	The sequence $\{\xn,\pmb{\Phi}_i^{(n)}\}$ satisfies
%	\begin{align}\label{a2_2}
%		&\hat{\varphi}^{(n)}(\mathbf{x}^{(n)},\mathbf{S}^{(n)})-\hat{\varphi}^{(n)}(\mathbf{x}^{(n+1)},\mathbf{S}^{(n+1)})\nonumber\\
%		&\geq \Big(\frac{\rho(f_0)+\rho(g_0)}{2}+\sum_{i=1}^I\tr(\pmb{\Phi}_i^{(n+1)})\frac{\rho(\pmb{\mathcal{F}}_i)}{2}\Big)\lVert\mathbf{x}^{(n+1)}-\mathbf{x}^{(n)}\rVert^2\nonumber\\
%		&\qquad+\sum_{i=1}^I\tr(\pmb{\Phi}_i^{(n+1)})\frac{\rho(\pmb{\mathcal{G}}_i)}{2}\lVert\mathbf{x}^{(n)}-\mathbf{x}^{(n-1)}\rVert^2,
%	\end{align}
%	where $\hat{\varphi}^{(n)}(\mathbf{x},\mathbf{S})=\varphi(\mathbf{x})++\sum_{i=1}^I\tau^{(n)}\tr(\mathbf{S}_i)$ and $\rho(\cdot)$ denotes the strong convexity parameter of the its corresponding function.
%\end{claim}

Since $\{\tau^{(n)}\}$ is bounded by Claim~\ref{claim1}, as indicated by the updating rule \eqref{sec4_alg1}, we have
\begin{equation}\label{a2_1}
	\tau^{(\tilde{n})}\geq\lambda_{\max}\big[\sum_{i=1}^I\pmb{\Phi}^{(\tilde{n}+1)}_i\big]+\delta_1.
\end{equation}
Invoking Weyl's inequality, \eqref{a2_1} further implies
\begin{equation}
	\tau^{(\tilde{n})}\geq\lambda_{\max}[\pmb{\Phi}^{(\tilde{n}+1)}_i]+\delta_1,~i\in\mathcal{I}.
\end{equation}
or equivalently, $\tau^{(n)}\mathbf{I}\succeq\pmb{\Phi}_i^{(n+1)}+\delta_1\mathbf{I}$ for all $n\geq\tilde{n}$. Then, in view of the complementary slackness \eqref{sec4_62}, it readily follows that $\mathbf{Z}_i^{(n+1)}\succ\mathbf{0}$. By \eqref{sec4_65}, we obtain $\mathbf{S}_i^{(n+1)}=\mathbf{0}$ for all $n\geq\tilde{n}$, which means that $\mathbf{x}^{(n)}$ is a feasible point of \eqref{sec4_3} for all $n\geq\tilde{n}$. Without loss of generality, considering a subsequence $\{\x^{(n_j)}\}$ of $\{\xn\}$, its limit point $\lim_{j\rightarrow\infty}\mathbf{x}^{(n_j)}=\bar{\mathbf{x}}$ is therefore feasible to \eqref{sec4_3}. Furthermore, \eqref{a2_1} implies that the subsequence $\{\pmb{\Phi}_i^{(n_j)}\}$ is bounded, and therefore we can assume that
\begin{equation}
	\lim_{j\rightarrow\infty}\pmb{\Phi}_i^{(n_j)}=\bar{\pmb{\Phi}}_i,~i\in\mathcal{I}.
\end{equation}

Now what remains to show is that any primal-dual pair of the limit point $(\bar{\mathbf{x}},\{\bar{\pmb{\Phi}}_i\})$ is a KKT stationary point of \eqref{sec4_3}. Let us  replace $n$ with $n_j$ in \eqref{sec4_6} and let $j\rightarrow\infty$. By noting that $\mathbf{x}^{(n_j)}$ and $\mathbf{x}^{(n_j+1)}$ are asymptotically close as indicated by Claim~\ref{claim2}, we obtain
\begin{subequations}\label{a2_3}
	\begin{align}
		\mathbf{0}\in\nabla f_0(\bar{\mathbf{x}})-\nabla g_0(\bar{\mathbf{x}})\qquad\qquad\qquad\qquad\qquad\qquad&\nonumber\\
		 +\sum_{i=1}^I\left(\bar{\pmb{\Phi}}_i*\left(\pmb{\mathcal{DF}}_i(\bar{\mathbf{x}})-\pmb{\mathcal{DG}}_i(\bar{\mathbf{x}})\right)\right)+\mathcal{N}(\Omega,\bar{\mathbf{x}})&\\
		\pmb{\mathcal{F}}_i(\bar{\mathbf{x}})-\pmb{\mathcal{G}}_i(\bar{\mathbf{x}})\preceq\mathbf{0},~\bar{\pmb{\Phi}}_i\succeq\mathbf{0},~i\in\mathcal{I}&\\
		\tr\left(\left(\pmb{\mathcal{F}}_i(\bar{\mathbf{x}})-\pmb{\mathcal{G}}_i(\bar{\mathbf{x}})\right)\bar{\pmb{\Phi}}_i\right)=0,~i\in\mathcal{I}&\\
		\bar{\mathbf{x}}\in\Omega,&
	\end{align}
\end{subequations}
which exactly represents the KKT conditions of the PSD-DC problem \eqref{sec4_3}. Noting the bounded nature of $\{\mathbf{x}^{(n)}\}$ assumed in A.2), it readily follows that there exists at least one limit point of $\{\mathbf{x}^{(n)}\}$ and by \eqref{a2_3}, any limit point of $\left\{\mathbf{x}^{(n)}\right\}$ is a KKT stationary point of \eqref{sec4_3}.

\section{Proof of Claim~\ref{claim1}}\label{appendix3}
We argue by contradiction. Assume the contrary, i.e., that $\lim_{n\rightarrow\infty}\tau^{(n)}=+\infty$.   From the updating rule \eqref{sec4_alg1}, it follows without loss of generality that there exist infinitely many indices $j$ so that
\begin{align}
	\tau^{(n_j)}&<\lambda_{\max}\Big[\sum_{i=1}^I\pmb{\Phi}_i^{(n_j+1)}\Big]+\delta_1,\\
	\tau^{(n_j)}&<\lVert\mathbf{x}^{(n_j+1)}-\mathbf{x}^{(n_j)}\rVert^{-1}.
\end{align}
By possibly restricting to a subsequence of $\left\{n_j\right\}$, without loss of generality, we can further assume that there exists at least some $i\in\mathcal{S}_{\mathcal{I}}$, where $\mathcal{S}_{\mathcal{I}}$ denotes a subset of $\mathcal{I}$, so that
\begin{align}\label{a3_1}
	&\lim_{j\rightarrow\infty}\lambda_{\max}[\pmb{\Phi}_i^{(n_j+1)}]=+\infty\nonumber\\
	&\qquad~~\Leftrightarrow\lim_{j\rightarrow\infty}\lVert\pmb{\Phi}_i^{(n_j+1)}\rVert_F=+\infty,~i\in\mathcal{S}_{\mathcal{I}}\\
	\label{a3_6}
	&\lim_{j\rightarrow\infty}\lVert\mathbf{x}^{(n_j+1)}-\mathbf{x}^{(n_j)}\rVert=0.
\end{align}
Let $\lim_{j\rightarrow\infty}\mathbf{x}^{(n_j)}=\bar{\x}$, and then we show that $\pmb{\mathcal{F}}_i(\bar{\x})-\pmb{\mathcal{G}}_i(\bar{\x})\nprec\mathbf{0}$ must hold for $i\in\mathcal{S}_{\mathcal{I}}$. Again, we show this by contradiction. If $\pmb{\mathcal{F}}_i(\bar{\x})-\pmb{\mathcal{G}}_i(\bar{\x})\prec\mathbf{0}$, then for a sufficiently large $j$, we must have:
\begin{multline}
	\pmb{\mathcal{F}}_i(\mathbf{x}^{(n_j+1)})-\pmb{\mathcal{G}}_i(\mathbf{x}^{(n_j)})\\
	-\pmb{\mathcal{DG}}(\mathbf{x}^{(n_j)})(\mathbf{x}^{(n_j+1)}-\mathbf{x}^{(n_j)})-\mathbf{S}_i^{(n_j+1)}\prec\mathbf{0}.
\end{multline}
This is due to \eqref{a3_6} such that
\begin{multline}
	\lim_{j\rightarrow\infty}\Big(\pmb{\mathcal{F}}_i(\mathbf{x}^{(n_j+1)})-\pmb{\mathcal{G}}_i(\mathbf{x}^{(n_j)})\\
	-\pmb{\mathcal{DG}}(\mathbf{x}^{(n_j)})(\mathbf{x}^{(n_j+1)}-\mathbf{x}^{(n_j)})\Big)
	=\pmb{\mathcal{F}}_i(\bar{\x})-\pmb{\mathcal{G}}_i(\bar{\x})
\end{multline}
as well as $\mathbf{S}_i^{(n_j+1)}\preceq\mathbf{0}$.
By the complementary slackness condition \eqref{sec4_64}, it readily follows that provided $j$ becomes sufficiently large, we have $\pmb{\Phi}_i^{(n_j+1)}=\mathbf{0}$ for $i\in\mathcal{S}_{\mathcal{I}}$, which contradicts the previous result of \eqref{a3_1}. Therefore, we have $\pmb{\mathcal{F}}_i(\bar{\x})-\pmb{\mathcal{G}}_i(\bar{\x})\nprec\mathbf{0}$ for $i\in\mathcal{S}_{\mathcal{I}}$.

Let us assume, without loss of generality, that
\begin{equation}\label{a3_7}
	\lim_{j\rightarrow\infty}\frac{\pmb{\Phi}_i^{(n_j+1)}}{\sum_{i=1}^I\lVert\pmb{\Phi}_i^{(n_j+1)}\rVert_F}=\hat{\pmb{\Phi}}_i\succeq\mathbf{0}.
\end{equation}
and it is easy to observe that $\hat{\pmb{\Phi}}_i=\mathbf{0}$ for $i\in\mathcal{I}\backslash\mathcal{S}_{\mathcal{I}}$ and $\hat{\pmb{\Phi}}_i\neq\mathbf{0}$ for $i\in\mathcal{S}_{\mathcal{I}}$. We now replace $n$ with $n_j$ in \eqref{sec4_6}. Dividing both sides of \eqref{sec4_61} by $\sum_{i=1}^I\lVert\pmb{\Phi}_i^{(n_j+1)}\rVert_F$, as well as taking the limit as  $j\rightarrow\infty$ and using the result of \eqref{a3_6}, we obtain
\begin{equation}\label{sec4_7}
	 \mathbf{0}\in\sum_{i\in\mathcal{S}_{\mathcal{I}}}\hat{\pmb{\Phi}}_i*\big(\pmb{\mathcal{DF}}_i(\bar{\mathbf{x}})-\pmb{\mathcal{DG}}_i(\bar{\mathbf{x}})\big)+\mathcal{N}(\Omega,\bar{\mathbf{x}}).
\end{equation}
%which is equivalent to
%\begin{equation}\label{a3_5}
%	\sum_{i\in\mathcal{S}_{\mathcal{I}}}\tr\left(\hat{\pmb{\Phi}}_i\frac{\partial \left(\pmb{\mathcal{F}}_i(\bar{\mathbf{x}})-\pmb{\mathcal{G}}_i(\bar{\mathbf{x}})\right)}{\partial x_j}\right)\geq 0,\quad j=1,\cdots,n,
%\end{equation}
%Note that \eqref{sec4_7} equivalently means that
%\begin{equation}\label{a3_3}
%	\tr\left(\hat{\pmb{\Phi}}_i\pmb{\Theta}_i(\bar{\mathbf{x}})(\mathbf{d})\right)\geq 0,~\forall \mathbf{d}
%\end{equation}

However, A.1) indicates that there exists some $\mathbf{h}\in\mathbb{C}^{n}$ so that $\left(\pmb{\mathcal{DF}}_i(\bar{\mathbf{x}})-\pmb{\mathcal{DG}}_i(\bar{\mathbf{x}})\right)\mathbf{h}\prec\mathbf{0}$ for $i\in\mathcal{S}_{\mathcal{I}}$. It is seen that \eqref{sec4_7} contradicts A.1).
%\begin{equation}\label{a3_4}
%	\tr\left(\hat{\pmb{\Phi}}_i\left(\pmb{\Theta}_i(\bar{\mathbf{x}})(\mathbf{h})\right)\right)\leq 0,~i\in\mathcal{S}_{\mathcal{I}},
%\end{equation}
Now we can assume that there exists a finite index $\tilde{n}$ so that
\begin{equation}\label{a3_5}
	\tau^{(n)}=\tau^{(\tilde{n})},~\forall n\geq\tilde{n}.
\end{equation}
\bibliographystyle{IEEEtran}
\bibliography{refs}

\begin{thebibliography}{10}
\providecommand{\url}[1]{#1}
\csname url@samestyle\endcsname
\providecommand{\newblock}{\relax}
\providecommand{\bibinfo}[2]{#2}
\providecommand{\BIBentrySTDinterwordspacing}{\spaceskip=0pt\relax}
\providecommand{\BIBentryALTinterwordstretchfactor}{4}
\providecommand{\BIBentryALTinterwordspacing}{\spaceskip=\fontdimen2\font plus
\BIBentryALTinterwordstretchfactor\fontdimen3\font minus
  \fontdimen4\font\relax}
\providecommand{\BIBforeignlanguage}[2]{{%
\expandafter\ifx\csname l@#1\endcsname\relax
\typeout{** WARNING: IEEEtran.bst: No hyphenation pattern has been}%
\typeout{** loaded for the language `#1'. Using the pattern for}%
\typeout{** the default language instead.}%
\else
\language=\csname l@#1\endcsname
\fi
#2}}
\providecommand{\BIBdecl}{\relax}
\BIBdecl

\bibitem{Mobile_threats}
``{M}obile cyber threats: {K}aspersky lab \& {INTERPOL} joint report,''
  {INTERPOL} and Kaspersky {L}ab, Tech. Rep., Oct. 2014.

\bibitem{WP}
C.~Timberg, ``German researchers discover a flaw that could let anyone listen
  to your cell calls,'' \emph{Washington Post: The Switch}, Dec. 2014,
  available online:
  https://www.washingtonpost.com/news/the-switch/wp/2014/12/18/german-researchers-discover-a-flaw-that-could-let-anyone-listen-to-your-cell-calls-and-read-your-texts/.

\bibitem{wyner}
A.~D. Wyner, ``The wire-tap channel,'' \emph{Bell Syst. Tech. J.}, vol.~54,
  no.~8, pp. 1355--1387, Jan. 1975.

\bibitem{Information_security}
Y.~Liang, H.~V. Poor, and S.~Shamai, ``Information theoretical security,''
  \emph{Found. Trends Commun. Inf. Theory}, vol.~5, no. 4-5, pp. 355--580,
  2008.

\bibitem{phy_security_multiantenna}
Y.-W.~P. Hong, P.-C. Lan, and C.-C.~J. Kuo, ``Enhancing physical-layer secrecy
  in multiantenna wireless systems,'' \emph{IEEE Signal Process. Mag.},
  vol.~30, no.~5, pp. 29--40, Sep. 2013.

\bibitem{phy_security_survey}
A.~Mukherjee, S.~A.~A. Fakoorian, J.~Huang, and A.~L. Swindlehurst,
  ``Principles of physical layer security in multiuser wireless networks: {A}
  survey,'' \emph{IEEE Commun. Surveys Tuts.}, vol.~16, no.~3, pp. 1550--1573,
  Third Quarter 2014.

\bibitem{Unified}
Y.~Rong, X.~Tang, and Y.~Hua, ``A unified framework for optimizing linear
  nonregenerative multicarrier {MIMO} relay communication systems,'' \emph{IEEE
  Trans. Signal Process.}, vol.~57, no.~12, pp. 4837--4851, Aug. 2009.

\bibitem{xing}
C.~Xing, S.~Ma, and Y.-C. Wu, ``Robust joint design of linear relay precoder
  and destination equalizer for dual-hop amplify-and-forward {MIMO} relay
  systems,'' \emph{IEEE Trans. Signal Process.}, vol.~58, no.~4, pp.
  2273--2283, Apr. 2010.

\bibitem{Lai}
L.~Lai and H.~E. Gamal, ``The relay eavesdropper channel: {C}ooperation for
  secrecy,'' \emph{IEEE Trans. Inf. Theory}, vol.~54, no.~9, pp. 4005--4019,
  Sep. 2008.

\bibitem{Secrecy_Cooperation}
H.-M. Wang and X.-G. Xia, ``Enhancing wireless secrecy via cooperation:
  {S}ignal design and optimization,'' \emph{IEEE Commun. Mag.}, vol.~53,
  no.~12, pp. 47--53, Dec. 2015.

\bibitem{Relay_Security_AFDF}
L.~Dong, Z.~Han, A.~P. Petropulu, and H.~V. Poor, ``Improving wireless physical
  layer security via cooperating relays,'' \emph{IEEE Trans. Signal Process.},
  vol.~58, no.~3, pp. 1875--1888, Mar. 2010.

\bibitem{Relay_Security_DF}
J.~Li, A.~P. Petropulu, and S.~Weber, ``On cooperative relaying schemes for
  wireless physical layer security,'' \emph{IEEE Trans. Signal Process.},
  vol.~59, no.~10, pp. 4985--4997, Oct. 2011.

\bibitem{Relay_Beamforming}
Y.~Yang, Q.~Li, W.-K. Ma, J.~Ge, and P.~C. Ching, ``Cooperative secure
  beamforming for {AF} relay networks with multiple eavesdroppers,'' \emph{IEEE
  Signal Process. Lett.}, vol.~20, no.~1, pp. 35--38, Jan. 2013.

\bibitem{6823730}
H.-M. Wang, F.~Liu, and X.-G. Xia, ``Joint source-relay precoding and power
  allocation for secure amplify-and-forward {MIMO} relay networks,'' \emph{IEEE
  Trans. Inf, Forensics Security}, vol.~9, no.~8, pp. 1240--1250, Aug. 2014.

\bibitem{6051523}
C.~Jeong, I.-M. Kim, and D.~I. Kim, ``Joint secure beamforming design at the
  source and the relay for an amplify-and-forward {MIMO} untrusted relay
  system,'' \emph{IEEE Trans. Signal Process.}, vol.~60, no.~1, pp. 310--325,
  Jan. 2012.

\bibitem{6655008}
S.~Vishwakarma and A.~Chockalingam, ``Amplify-and-forward relay beamforming for
  secrecy with cooperative jamming and imperfect {CSI},'' in \emph{Proc. IEEE
  Int. Conf. Commun. (ICC), 2013}, Budapest, Hungry, Jun. 2013, pp. 3047--3052.

\bibitem{6956810}
C.~Zhang, H.~Gao, H.~Liu, and T.~Lv, ``Robust beamforming and jamming for
  secure af relay networks with multiple eavesdroppers,'' in \emph{Proc. IEEE
  Military Commun. Conf. (MILCOM), 2014}, Baltimore, MD, Oct. 2014, pp.
  495--500.

\bibitem{Robust_Secure_Relaying}
X.~Wang, K.~Wang, and X.-D. Zhang, ``Secure relay beamforming with imperfect
  channel side information,'' \emph{IEEE Trans. Veh. Technol.}, vol.~62, no.~5,
  pp. 2140--2155, Jun. 2013.

\bibitem{Robust_AF_AN}
Q.~Li, Y.~Yang, W.-K. Ma, M.~Lin, J.~Ge, and J.~Lin, ``Robust cooperative
  beamforming and artificial noise design for physical-layer secrecy in {AF}
  multi-antenna multi-relay networks,'' \emph{IEEE Trans. Signal Process.},
  vol.~63, no.~1, pp. 206--220, Jan. 2015.

\bibitem{6819004}
K.~Jayasinghe, P.~Jayasinghe, N.~Rajatheva, and M.~{Latva-aho}, ``Secure
  beamforming design for physical layer network coding based {MIMO} two-way
  relaying,'' \emph{IEEE Commun. Lett.}, vol.~18, no.~7, pp. 1270--1273, Jul.
  2014.

\bibitem{6560019}
M.~Zhang, J.~Huang, H.~Yu, H.~Luo, and W.~Chen, ``{QoS}-based source and relay
  secure optimization design with presence of channel uncertainty,'' \emph{IEEE
  Commun. Lett.}, vol.~17, no.~8, pp. 1544--1547, Aug. 2013.

\bibitem{Ding}
Z.~Chu, K.~Cumanan, M.~Xu, and Z.~Ding, ``Robust secrecy rate optimisations for
  multiuser multiple-input-single-output channel with device-to-device
  communications,'' \emph{IET Commun.}, vol.~9, no.~9, pp. 396--403, Feb. 2015.

\bibitem{D2D}
{3GPP~TR~36.843~V~12.0.1}, ``3rd generation partnership project; technical
  specification group radio access network; study on {LTE} device to device
  proximity services; radio aspects ({R}elease 12),'' Mar. 2014.

\bibitem{SDR}
Z.-Q. Luo, W.-K. Ma, A.~M.-C. So, Y.~Ye, and S.~Zhang, ``Semidefinite
  relaxation of quadratic optimization problems,'' \emph{IEEE Signal Process.
  Mag.}, vol.~27, no.~3, pp. 20--34, May 2010.

\bibitem{SRT}
Y.~Zou, X.~Wang, W.~Shen, and L.~Hanzo, ``Security versus reliability analysis
  of opportunistic relaying,'' \emph{IEEE Trans. Veh. Technol.}, vol.~63,
  no.~6, pp. 2653--2661, Jul. 2014.

\bibitem{Yulong_Survey}
Y.~Zou, X.~Wang, and L.~Hanzo, ``A survey on wireless security: {T}echnical
  challenges, recent advances and future trends,'' \emph{Proc. IEEE}, 2015, to
  be published. [Online] Available: http://arxiv.org/abs/1505.07919.

\bibitem{CCP}
B.~K. Sriperumbudur and G.~R.~G. Lanckriet, ``On the convergence of the
  concave-convex procedure,'' \emph{Advances Neural Inf. Process. Syst. 22},
  pp. 1759--1767, 2009.

\bibitem{shannon}
C.~E. Shannon, ``A mathematical theory of communications,'' \emph{Bell Syst.
  Tech. J.}, vol.~27, pp. 379--423, 1948.

\bibitem{hero}
A.~O. Hero, ``Secure space-time communication,'' \emph{IEEE Trans. Inf.
  Theory}, vol.~49, no.~12, pp. 3235--3249, Dec. 2003.

\bibitem{Imperfect_CSI}
A.~Pascual-Iserte, D.~P. Palomar, A.~I. Perez-Neira, and M.~A. Lagunas, ``A
  robust maximin approach for {MIMO} communications with imperfect channel
  state information based on convex optimization,'' \emph{IEEE Trans. Signal
  Process.}, vol.~54, no.~1, pp. 346--360, Jan. 2006.

\bibitem{Robust_Rong}
Y.~Rong, ``Robust design for linear non-regenerative {MIMO} relays with
  imperfect channel state information,'' \emph{IEEE Trans. Signal Process.},
  vol.~59, no.~5, pp. 2455--2460, Feb. 2011.

\bibitem{Yang_TVT}
J.~Yang, B.~Champagne, Y.~Zou, and L.~Hanzo, ``Joint optimization of
  transceiver matrices for {MIMO}-aided multiuser {AF} relay networks:
  {I}mproving the {QoS} in the presence of {CSI} errors,'' \emph{IEEE Trans.
  Veh. Technol.}, 2015, {IEEE Xplore early access}.

\bibitem{Yang_ICC}
------, ``{MIMO} {AF} relaying security: {R}obust transceiver design in the
  presence of multiple eavesdroppers,'' in \emph{Proc. IEEE Int. Conf. Commun.
  (ICC), 2015}, London, U.K., Jun. 2015, pp. 4937--4942.

\bibitem{Convex_Optimization}
S.~Boyd and L.~Vandenberghe, \emph{Convex Optimization}.\hskip 1em plus 0.5em
  minus 0.4em\relax Cambridge, U.K.: Cambridge Univ. Press, 2004.

\bibitem{Charnes_Cooper}
A.~Charnes and W.~W. Cooper, ``Programming with linear fractional
  functionals,'' \emph{Naval Res. Logist. Quart.}, vol.~9, no. 3/4, pp.
  181--186, Dec. 1962.

\bibitem{SeDuMi}
J.~F. Sturm, ``Using {S}e{D}u{M}i 1.02, a {MATLAB} toolbox for optimization
  over symmetric cones,'' \emph{Optim. Methods Softw.}, vol.~11, no. 1-4, pp.
  625--653, Jan. 1999.

\bibitem{mosek}
E.~D. Andersen and K.~D. Andersen, ``{MOSEK} modeling manual,''
  \url{http://mosek.com}, Aug. 2013.

\bibitem{Interior_Point}
Y.~Nesterov and A.~Nemirovski, \emph{Interior Point Polynomial Time Methods in
  Convex Programming: Theory and Applications}.\hskip 1em plus 0.5em minus
  0.4em\relax Philadelphia, PA: SIAM, 1994.

\bibitem{SDP}
L.~Vandenberghe and S.~Boyd, ``Semidefinite programming,'' \emph{SIAM Rev.},
  vol.~38, no.~1, pp. 49--95, Mar. 1996.

\bibitem{DC_Overview}
R.~Horst and N.~V. Thoai, ``{DC} programming: {O}verview,'' \emph{J. Optim.
  Theory Appl.}, vol. 103, no.~1, pp. 1--43, Oct. 1999.

\bibitem{Penalty_Nonlinear}
S.-P. Han and O.~L. Mangasarian, ``Exact penalty functions in nonlinear
  programming,'' \emph{Math. Programming}, vol.~17, no.~1, pp. 251--269, 1979.

\bibitem{Penalty_Constrained}
G.~{Di~Pillo} and L.~Grippo, ``Exact penalty functions in constrained
  optimization,'' \emph{SIAM J. Control Optim.}, vol.~26, no.~6, pp.
  1333--1360, Nov. 1989.

\bibitem{SOCP}
M.~S. Lobo, L.~Vandenberghe, S.~Boyd, and H.~Lebret, ``Applications of
  second-order cone programming,'' \emph{Linear Algebra Appl.}, vol. 284, no.
  1-3, pp. 193--228, Nov. 1998.

\end{thebibliography}
%\bibliography{\detokenize{/Users/carteryang/Dropbox/Bibliographies/refs}}
%\bibliography{\detokenize{/Users/$USER/Dropbox/Bibliographies/refs}}
%
%% biography section
%%
%% If you have an EPS/PDF photo (graphicx package needed) extra braces are
%% needed around the contents of the optional argument to biography to prevent
%% the LaTeX parser from getting confused when it sees the complicated
%% \includegraphics command within an optional argument. (You could create
%% your own custom macro containing the \includegraphics command to make things
%% simpler here.)
%%\begin{biography}[{\includegraphics[width=1in,height=1.25in,clip,keepaspectratio]{mshell}}]{Michael Shell}
%% or if you just want to reserve a space for a photo:
%\end{IEEEbiography}
%% insert where needed to balance the two columns on the last page with
%%\newpage
%% You can push biographies down or up by placing
%% a \vfill before or after them. The appropriate
%% use of \vfill depends on what kind of text is
%% on the last page and whether or not the columns
%% are being equalized.
%
%%\vfill
%\newpage
%% Can be used to pull up biographies so that the bottom of the last one
%% is flush with the other column.
%%\enlargethispage{-5in}

\end{document}